\documentclass[]{fairmeta}
\usepackage{xcolor}   
\usepackage{microtype}
\usepackage{graphicx}
\usepackage{booktabs} 

\usepackage{comment}
\usepackage{amsmath}
\usepackage{amssymb}
\usepackage{mathtools}
\usepackage{amsthm}
\usepackage{bm}
\usepackage{tcolorbox}
\hypersetup{
    colorlinks=true,
    citecolor =cyan,
    linkcolor=magenta,
    urlcolor=blue,
}

\usepackage{microtype}      
\usepackage{graphicx}
\usepackage{booktabs}       
\usepackage[ruled,algo2e]{algorithm2e}
\usepackage{algorithm}
\usepackage{comment}
\usepackage{nicefrac} 

\usepackage{amsmath}
\usepackage{amssymb}
\usepackage{mathtools}
\usepackage{amsthm}
\usepackage{comment}
\usepackage{cleveref}
\usepackage{minitoc}
\theoremstyle{plain}

\theoremstyle{definition}

\theoremstyle{remark}

\usepackage[textsize=tiny]{todonotes}
\usepackage[labelfont=it,font+=smaller]{caption}
\usepackage[labelfont=it]{subcaption}
\usepackage{makecell}
\usepackage{arydshln}
\usepackage{multirow}

\definecolor{lightblue}{rgb}{0.7725490196, 0.89019607843, 0.9294117647}
\definecolor{darkblue}{rgb}{0.11764705882, 0.11764705882, 0.54509803921}
\usepackage{rotating}

\usepackage{xspace}
\usepackage{bbm}
\input{mathlig}
\usepackage{mathtools}
\usepackage{relsize}
\usepackage{mathrsfs}
\usepackage{dsfont}
\DeclarePairedDelimiterX{\inp}[2]{\langle}{\rangle}{#1, #2}
\makeatletter
\newcommand*\bigcdot{\mathpalette\bigcdot@{.5}}
\newcommand*\bigcdot@[2]{\mathbin{\vcenter{\hbox{\scalebox{#2}{$\m@th#1\bullet$}}}}}
\makeatother

\newcommand{\muspace}{\mspace{1mu}}

\DeclareRobustCommand{\scond}{\mathchoice{\muspace\vert\muspace}{\vert}{\vert}{\vert}}
\mathlig{|}{\scond}

\DeclareRobustCommand{\discint}{\mathchoice{\mspace{-1.5mu}:\mspace{-1.5mu}}{\mspace{-1.5mu}:\mspace{-1.5mu}}{:}{:}}
\mathlig{::}{\discint}

\newcommand{\Xv}{{\bf X}}
\newcommand{\Yv}{{\bf Y}}

\newcommand{\av}{{\bf a}}
\newcommand{\bv}{{\bf b}}

\newcommand{\xv}{{\bf x}}
\newcommand{\yv}{{\bf y}}

\newcommand{\Xh}{{\hat{X}}}

\def\th{\theta}

\def\textiid{i.i.d.\@\xspace}
\newcommand\iid{\ifmmode\text{ i.i.d. } \else \textiid \fi}

\def\mathllap{\mathpalette\mathllapinternal}
\def\mathllapinternal#1#2{%
  \llap{$\mathsurround=0pt#1{#2}$}}

\def\clap#1{\hbox to 0pt{\hss#1\hss}}
\def\mathclap{\mathpalette\mathclapinternal}
\def\mathclapinternal#1#2{%
  \clap{$\mathsurround=0pt#1{#2}$}}

\let\oldstackrel\stackrel
\renewcommand{\stackrel}[2]{\oldstackrel{\mathclap{#1}}{#2}}

\DeclarePairedDelimiterX{\infdivx}[2]{(}{)}{%
  #1\;\delimsize\|\;#2%
}

\renewcommand{\hbar}{h\mathllap{\overline{\vphantom{h}\hphantom{\rule{4.6pt}{0pt}}}\mspace{0.77mu}}}

\catcode`~=11 
\newcommand{\urltilde}{\kern -.06em\lower -.06em\hbox{~}\kern .02em}
\catcode`~=13 

\DeclarePairedDelimiterX{\norm}[1]{\lVert}{\rVert}{#1}
\DeclarePairedDelimiterX{\abs}[1]{\lvert}{\rvert}{#1}

\usepackage{xparse}

\let\oldpartial\partial
\renewcommand*{\partial}{\mathop{}\!\oldpartial}
\colorlet{tablerowcolor}{lightblue}

\usepackage{enumitem}

\usepackage{dashrule}

\usepackage{tcolorbox}\tcbuselibrary{skins}
\tcbuselibrary{breakable}

\tcbset{gray_box/.style={
    enhanced,
    breakable,
    size=fbox,
    boxrule=2pt,
    arc=2mm,
    auto outer arc,
    left=1pt,
    right=1pt,
    top=1pt,
    bottom=1pt,
}}

\tcbset{red_box/.style={
    enhanced,
    size=fbox,
    boxrule=2pt,
    arc=2mm,
    auto outer arc,
    left=1pt,
    right=1pt,
    top=1pt,
    bottom=1pt,
    colback=red!10,
    colframe=red!75!black,
    coltitle=white, 
}}

\tcbset{
    brown_box/.style={
        enhanced,
        colback=brown!10,
        colframe=brown!80!black,
        coltitle=white,
        arc=2mm,
        auto outer arc,
        left=1pt,
        right=1pt,
        top=1pt,
        bottom=1pt,
        boxrule=2pt,
    }
}

\definecolor{green_color}{RGB}{0, 150, 0}
\tcbset{green_box/.style={
    enhanced,
    breakable,
    size=fbox,
    boxrule=2pt,
    arc=2mm,
    auto outer arc,
    left=1pt,
    right=1pt,
    top=1pt,
    bottom=1pt,
    colback=green!8,
    colframe=green_color,
    coltitle=white, 
}}

\definecolor{blue_color}{RGB}{8, 104, 172}
\tcbset{blue_box/.style={
    enhanced,
    breakable,
    size=fbox,
    boxrule=2pt,
    arc=2mm,
    auto outer arc,
    left=1pt,
    right=1pt,
    top=1pt,
    bottom=1pt,
    colback=blue!5,
    colframe=blue_color,
    coltitle=white, 
}}

\RenewDocumentCommand\eqref{D<>{Eq.}om}{%
\IfNoValueTF{#2}
{#1~\oldeqref{#3}}
{(#2 #1~\textup{\ref{#3}})}%
}

\usepackage{scrwfile}
\TOCclone[\appendixname]{toc}{atoc}
\newcommand\StartAppendixEntries{}
\AfterTOCHead[toc]{%
  \renewcommand\StartAppendixEntries{\value{tocdepth}=-10000\relax}%
}
\AfterTOCHead[atoc]{%
  \edef\maintocdepth{\the\value{tocdepth}}%
  \value{tocdepth}=-10000\relax%
  \renewcommand\StartAppendixEntries{\value{tocdepth}=\maintocdepth\relax}%
}

\usepackage{booktabs}
\usepackage{siunitx}
\usepackage{fontspec}
\defaultfontfeatures{Ligatures=TeX}

\newfontfamily\ipafont[
  Path=Fonts/,
  UprightFont=Charis-Regular.ttf,
  BoldFont=Charis-Bold.ttf,
  ItalicFont=Charis-Italic.ttf,
  BoldItalicFont=Charis-BoldItalic.ttf
]{Charis-Regular.ttf}

\newcommand{\IPA}[1]{{\ipafont #1}}

\makeatletter
\@ifundefined{BreakableUnderscore}{}{%
  
}
\makeatother
\usepackage[strings]{underscore}

\title{GSRM: Generative Speech Reward Model for Speech RLHF}

\author[1,2,*,\dagger]{Maohao Shen}
\author[1,*]{Tejas Jayashankar}
\author[1,*]{Osama Hanna}
\author[1,*]{Naoyuki Kanda}

\author[1,3,\dagger]{\\Yancheng Wang}
\author[1]{Kateřina Žmolíková}
\author[1]{Ruiming Xie}
\author[1]{Niko Moritz}
\author[1,4,\dagger]{Anfeng Xu}
\author[1]{Yashesh Gaur}

\author[2]{Gregory Wornell}
\author[1]{Qing He}
\author[1]{Jilong Wu}

\affiliation[1]{Meta Superintelligence Labs}
\affiliation[2]{Massachusetts Institute of Technology}
\affiliation[3]{Arizona State University}
\affiliation[4]{University of Southern California}

\contribution[*]{Core contributors}
\contribution[\dagger]{Work done during internship at Meta Superintelligence Labs}

\correspondence{Maohao Shen at \email{maohao@mit.edu}}
\date{\today}

\begin{document}
\abstract{
Recent advances in speech language models, such as GPT-4o Voice Mode and Gemini Live, have demonstrated promising speech generation capabilities. Nevertheless, the aesthetic naturalness of the synthesized audio still lags behind that of human speech. Enhancing generation quality requires a reliable evaluator of speech naturalness. However, existing naturalness evaluators typically regress raw audio to scalar scores, offering limited interpretability of the evaluation and moreover fail to generalize to speech across different taxonomies. Inspired by recent advances in generative reward modeling, we propose the \emph{Generative Speech Reward Model} (GSRM), a reasoning-centric reward model tailored for speech. The GSRM is trained to decompose speech naturalness evaluation into an interpretable acoustic feature extraction stage followed by feature-grounded chain-of-thought reasoning, enabling explainable judgments. To achieve this, we curated a large-scale human feedback dataset comprising 31k expert ratings and an out-of-domain benchmark of real-world user-assistant speech interactions. Experiments show that GSRM substantially outperforms existing speech naturalness predictors, achieving model–human correlation of naturalness score prediction that approaches human inter-rater consistency. We further show how GSRM can improve the naturalness of speech LLM generations by serving as an effective verifier for online RLHF. 
}

\maketitle

\newcommand{\bluet}[1]{\textcolor{blue}{#1}}
\renewcommand{\bluet}[1]{#1}
\newcommand{\greent}{\textcolor{Green}}

\newcommand{\Dir}{\mathsf{Dir}}
\newcommand{\pout}{p_{\textsf{ood}}}

\renewcommand{\av}{\boldsymbol{\alpha}}

\newcommand{\piv}{\boldsymbol{\pi}}
\newcommand{\Piv}{\boldsymbol{\Pi}}

\renewcommand{\xv}{\boldsymbol{\rm x}}
\renewcommand{\Xv}{\boldsymbol{\rm X}}
\renewcommand{\yv}{\boldsymbol{\rm y}}
\renewcommand{\Yv}{\boldsymbol{\rm Y}}
\renewcommand{\Xh}{\boldsymbol{\rm \hat{X}}}
\newcommand{\vol}{\mathsf{vol}}
\newcommand{\green}[1]{\textcolor{ForestGreen}{#1}}
\newcommand{\red}[1]{\textcolor{red}{#1}}
\newcommand{\blue}[1]{\textcolor{blue}{#1}}
\newcommand{\brown}[1]{\textcolor{brown}{#1}}
\newcommand{\orange}[1]{\textcolor{orange}{#1}}
\newcommand{\purple}[1]{\textcolor{purple}{#1}}

\newcommand{\pbar}{\phi}
\newcommand{\pnoise}{\textcolor{red}{p_{\mathsf{n}}}}
\newcommand{\pdata}{\green{p_{\mathsf{d}}}}
\newcommand{\xdata}{\green{x^N}}
\newcommand{\pext}{\textcolor{blue}{\phi_{\th}}}

\newcommand{\etab}{\boldsymbol{\eta}}
\newcommand{\lambdab}{\boldsymbol{\lambda}}
\newcommand{\Poi}{\mathsf{Poi}}
\newcommand{\GammaDist}{\mathsf{G}}

\renewcommand{\av}{\boldsymbol{\alpha}}
\renewcommand{\bv}{\boldsymbol{\beta}}
\newcommand{\ev}{{\bf e}}
\newcommand{\wv}{\mathbf{w}}
\newcommand{\const}{\mathsf{(const.)}}

\newcommand{\catent}{H}
\newcommand{\diffent}{h}
\newcommand{\mi}{I}

\newcommand{\zwh}[1]{\textcolor{red}{ZWH: #1}}

\section{Introduction} \label{sec:intro}
Despite rapid progress in speech language models, such as GPT-4o Voice Mode~\cite{hurst2024gpt} and Gemini Live~\cite{comanici2025gemini}, the naturalness of synthesized speech still lags behind human audio, with noticeable deficits in human-likeness, prosody, and emotional nuance, thus limiting the level of human-like engagement in real-worl applications. Bridging this gap requires a reliable evaluator of speech naturalness to guide the improvement of speech generation systems.

Traditional evaluators, including MOS predictors~\citep{lo2019mosnet, saeki2022utmos, baba2024utmosv2, wang2024uncertainty, samos2024, hoq2025fusemos}, enable scalable speech evaluation but ultimately reduce perceptual judgments to a single scalar score. While effective for large-scale benchmarking, such metrics provide limited insight into \emph{why} a given speech sample is perceived as unnatural. Speech naturalness is inherently multi-dimensional, including factors such as expressiveness, intonation, and pacing. Forcing a single regressor to capture all these aspects often leads to poor generalization~\citep{lo2019mosnet, saeki2022utmos}. As a result, scalar predictors can yield unreliable reward signals when used for online reinforcement learning. In the broader text-based LLM literature, \emph{Generative Reward Models} (GRMs) have recently emerged as a promising alternative to scalar reward predictors. Rather than producing only a numerical score, GRMs generate explicit reasoning traces alongside their evaluations, more closely aligning with how humans justify judgments~\citep{mahan2024generative, liu2025inference, zhang2024generativeverifiers, liang2025grmcriteria}.

However, generative reward modeling remains relatively under-explored in the speech domain. AudioJudge~\cite{manakul2025audiojudge} evaluates the ability of frontier speech LLMs to judge audio quality and paralinguistic attributes through prompt engineering, revealing that most models struggle to produce reliable judgments. WavReward~\cite{ji2025wavreward} finetunes a GRM to assess specific speech attributes such as age, gender, pitch, and volume. Concurrent work likeSpeechJudge~\cite{zhang2025speechjudge} extends generative reward modeling to judge speech naturalness. While both SpeechJudge and our approach aim to construct synthetic CoT reasoning, they differ fundamentally in how the CoT traces are produced. SpeechJudge synthesizes CoT using a teacher speech LLM directly conditioned on the raw audio signal, resulting in reasoning that is mediated by implicit acoustic representations the speech LLM can capture. In contrast, we explicitly extract interpretable paralinguistic acoustic cues via signal processing and generate CoT using a text LLM conditioned on these features. This explicit grounding enables systematic ablation of which acoustic factors influence the predicted naturalness.  Moreover, although GRMs have been used to improve text-to-speech (TTS) systems~\cite{zhang2025speechjudge}, no prior work has leveraged these models to improve the naturalness of a full-duplex speech LLM via online reinforcement learning.

To this end, we introduce \textbf{GSRM}, a \emph{Generative Speech Reward Model} tailored for speech RLHF, with a focus on speech naturalness evaluation. GSRM explicitly decomposes speech judgment into two stages: (1) extracting interpretable paralinguistic acoustic features from raw audio, and (2) generating feature-grounded CoT reasoning traces to produce final evaluations. We summarize our main contributions as follows:
\begin{enumerate}[leftmargin=*]
    \item \textbf{High-quality Human Feedback Data.} In Section~\ref{sec:human_data}, we curate a large-scale human feedback dataset for speech naturalness, comprising 6.5K unique audio samples generated by controllable TTS systems. In addition, we collect 490 dialogue samples generated by an internal speech LLM to emulate real-world interactive scenarios.

    \item \textbf{Acoustic Feature–grounded Reasoning.} Unlike existing approaches, GSRM explicitly extracts paralinguistic acoustic features from speech and reasons over this evidence to judge audio quality. Section~\ref{sec:gsrm} introduces a scalable CoT synthesis pipeline that enables the generation of high-quality, feature-grounded reasoning trajectories.

    \item \textbf{Strong Empirical Performance.} 
    GSRM achieves human-level performance and
    substantially outperforming exisiting methods on naturalness prediction. Moreover, we take a first step toward online RLHF for speech LLMs. In Section~\ref{sec:rlhf}, by leveraging GSRM as a verifier, we successfully improve the naturalness of a speech LLM through online reinforcement learning.
\end{enumerate}

\section{Human Feedback Data for Speech Naturalness} \label{sec:human_data}
\begin{table*}[!t]
\centering
\footnotesize
\renewcommand{\arraystretch}{1}  
\setlength{\tabcolsep}{9pt}
\captionsetup{font=small}

\caption{\textbf{Annotation Rubric for Speech Naturalness.}
Each dialog is evaluated along multiple sub-dimensions that capture different aspects of conversational speech quality.}
\vspace{-0.6em}

\begin{tabular}{p{3.2cm} p{12.2cm}}
\toprule
\textbf{Sub-metric} & \textbf{Description} \\
\midrule
\textbf{Expressive Intensity} &
The degree of expressiveness, reflecting how emotionally vivid or subdued the speech sounds. \\

\textbf{Expressive Correctness} &
The appropriateness of the expressiveness w.r.t. the conversational context and semantic content. \\

\textbf{Intonation} &
The quality of intonation, including pitch variation, stress patterns, and overall prosodic naturalness. \\

\textbf{NSVS}&
The naturalness of non-speech vocalizations and filler words (e.g., breaths, pauses, hesitations). \\

\textbf{Mispronunciation} &
Whether mispronunciations are present in the speech, reflecting pronunciation accuracy. \\

\textbf{Pacing} &
The naturalness of the pacing or speaking rate, including timing, pauses, and temporal flow. \\

\midrule
\textbf{Human-likeness} &
An overall rating of how closely the speech resembles natural human conversational speech. \\
\bottomrule
\end{tabular}
\label{table:rubric}
\vspace{-1em}
\end{table*}

To reliably conduct speech naturalness assessment, we collect high-quality human annotations that capture both fine-grained attributes and overall human-likeness of conversational speech. These annotations serve as the foundation for training speech naturalness predictor.

\textbf{Data Annotation Rubric.}
To assess speech naturalness, we design a structured annotation rubric that decomposes naturalness into multiple relevant sub-metrics together with an overall \emph{human-likeness} score. Human raters evaluate each system response independently. All sub-metrics are rated on a 5-point Likert scale, where higher values indicate better quality. One exception is for mispronunciation, which is annotated using a categorical scale: 1 means a mispronunciation is present, 2 represents no mispronunciation, 3 means not sure. The details are provided in Table~\ref{table:rubric}.

\textbf{Conversational TTS Dataset (ConvTTS).}  
ConvTTS dataset is collected using an in-house TTS system capable of producing two-channel dialogues. The model generates synthetic dialogue between a user and a system by conditioning on reference speech examples along with textual instructions. The resulting datasets contains 6,579 dialogue samples, each of length 49.4 $\pm$19.3 seconds. We partition the samples into 4,579 training samples, 1,000 development samples, and 1,000 test samples, corresponding to 62.8, 13.7, and 13.8 hours of audio, respectively. Each sample is annotated for audio naturalness by human evaluators at the dialogue level. A subset of the samples also include an additional rating of the assistant speech reponse, totaling to 3,263 training, 711 development, and 717 test samples. On average, each audio recording is labeled by 6.8 annotators, with the majority of samples evaluated by at least five different human raters.

\textbf{Full-duplex Conversational Dataset (FDX-Conv).} 
FDX-Conv dataset is collected from an in-house speech LLM designed for full-duplex interaction, i.e., the system can listen and speak at the same time, allowing overlapping user and system speech rather than strictly turn-based exchanges. The dataset contains 490 dialogue samples, each with an averaged duration of 41.1 $\pm$ 19.0 seconds. Similarly, each sample has been rated by no fewer than five human raters. Since the speech is generated by a model distinct from the ConvTTS system, we treat FDX-Conv as an out-of-domain (OOD) benchmark for assessing the generalization capability of naturalness predictors.

\textbf{Inter-Rater Consistency of Human-Likeness Ratings.}
To ensure the reliability of the collected annotations, we measure inter-rater consistency for overall human-likeness ratings. Specifically, for each audio sample, we randomly partition the set of annotators into two disjoint subsets, compute the average rating within each subset, and measure the Pearson correlation coefficient between the two averages. The Pearson correlation coefficients are $0.533$ for ConvTTS and $0.532$ for FDX-Conv (see Figure~\ref{fig:human_pcc} in Appendix), indicating strong inter-rater agreement. These results also serve as an empirical upper bound on the consistency between an automatic naturalness predictor and human raters. We further analyze how individual sub-metrics relate to the overall human-likeness score in Appendix~\ref{sec:data_analysis}.
\section{Pilot Studies for Naturalness Prediction} \label{sec:pilot_studies}
In this section, we conduct pilot studies to assess the limitations of existing modeling paradigms for predicting speech naturalness. More details are included in Appendix~\ref{sec:app-setting}.
\begin{figure*}[!t]
    \centering
    \includegraphics[width=1.0\textwidth]{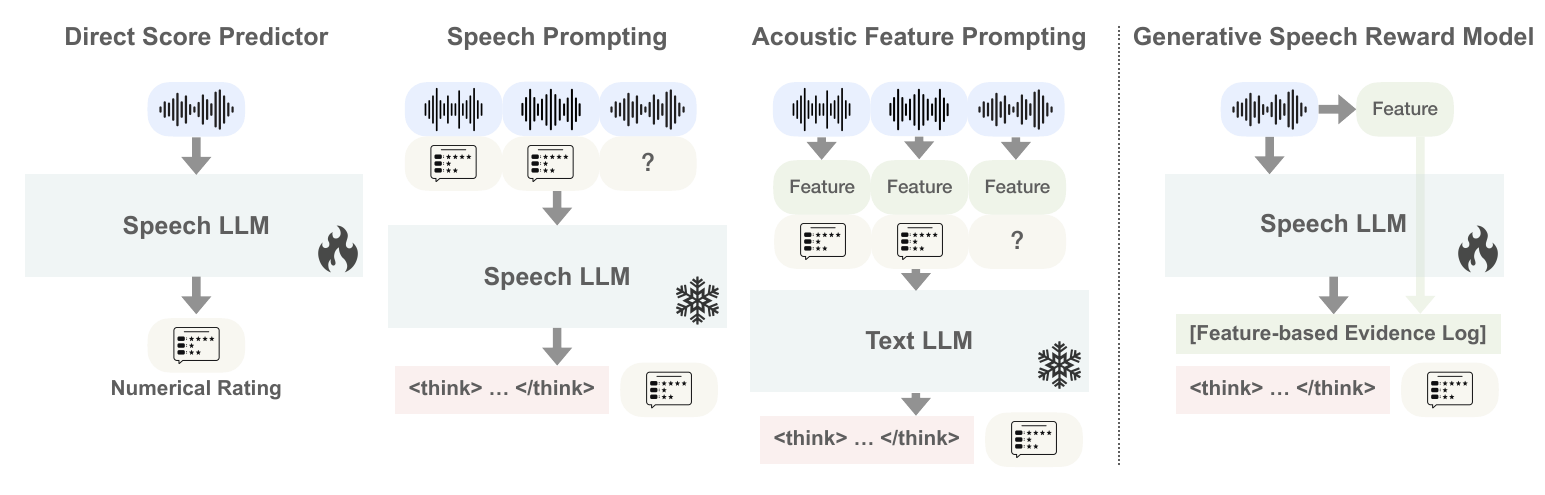}
\vspace{-1.5em}
\caption{\textbf{Comparison of Naturalness Predictors.} Generative Speech Reward Model (GSRM) integrates explicit acoustic feature extraction with feature-grounded chain-of-thought reasoning. Unlike other approaches, GSRM produces an interpretable evidence log derived from raw audio before reasoning or judging final numerical ratings.}
\label{fig:pilot}
\end{figure*}
\begin{table*}[!t]
\centering
\footnotesize
\captionsetup{font=small}
\caption{\textbf{Pilot Study Results.} Performance of baseline and frontier LLM-based predictors on speech naturalness evaluation. Correlation metrics are higher-is-better ($\uparrow$), while MSE is lower-is-better ($\downarrow$).}

\vspace{-0.6em}
\setlength{\tabcolsep}{4.8pt}
\begin{tabular}{l c
                S[table-format=1.3] S[table-format=1.3] S[table-format=1.3]
                S[table-format=1.3] S[table-format=1.3] S[table-format=1.3]}
\toprule
& & \multicolumn{3}{c}{\textbf{Validation}} & \multicolumn{3}{c}{\textbf{Out-of-domain}} \\
\cmidrule(lr){3-5}\cmidrule(lr){6-8}
\textbf{Method}
& \textbf{Input}
& {\textbf{Pearson} $\uparrow$}
& {\textbf{Spearman} $\uparrow$}
& {\textbf{MSE} $\downarrow$}
& {\textbf{Pearson} $\uparrow$}
& {\textbf{Spearman} $\uparrow$}
& {\textbf{MSE} $\downarrow$} \\
\midrule
Human Inter-Rater Consistency (Oracle)
& Speech
& 0.533 & 0.517 & 0.301
& 0.532 & 0.489 & 0.336 \\

\midrule
Regression-based Predictor
& Speech
& 0.215 & 0.170 & 0.433
& 0.252 & 0.226 & 0.357 \\

Gemini-2.5-Pro Speech Prompting
& Speech
& -0.050 & -0.078 & 1.009
&  0.140 &  0.136 & 0.531 \\

Gemini-2.5-Pro Acoustic Features Prompting
& Text
& 0.133 & 0.130 & 1.180
& 0.193 & 0.157 & 0.814 \\
\bottomrule
\end{tabular}
\label{table:pilot-results}
\vspace{-1em}
\end{table*}

\textbf{Direct Score Predictor.}
A natural baseline for speech naturalness prediction is to finetune a speech LLM to perform \emph{direct score prediction}. The model takes a speech recording as input and outputs numerical ratings for individual sub-metrics, followed by an overall human-likeness score (e.g., \texttt{human\_likeness: 3.0}), without generating intermediate reasoning or chain-of-thought (CoT). We implement this baseline using Qwen2.5-Omni-7B~\cite{xu2025qwen2}. The model is fine-tuned on the ConvTTS training set and evaluated on both the ConvTTS test set and the FDX-Conv OOD set. As shown in Table~\ref{table:pilot-results}, this approach achieves Pearson correlation coefficients of $0.215$ on the test set and $0.252$ on the OOD set. While these results indicate that the model captures some signal correlated with human judgments, a substantial gap remains compared to human inter-rater consistency, highlighting the limitations of direct score prediction without explicit reasoning.

\textbf{Frontier Speech Models Few-shot Prompting.}
Given the impressive speech generation capabilities of Gemini Live~\cite{comanici2025gemini}, we next examine whether it can be directly leveraged to judge speech naturalness via prompting. Prior work~\cite{manakul2025audiojudge} suggests that frontier speech LLMs often struggle to identify paralinguistic cues under zero-shot prompting, but that audio concatenation and in-context learning (ICL) examples can offer improvements. Following \citet{manakul2025audiojudge}, we randomly select five audio samples with human-likeness ratings spanning 1 to 5 as ICL examples. The corresponding audio clips and their ground-truth ratings are concatenated and provided as few-shot context to Gemini-2.5-Pro in voice mode. We then evaluate the model’s predictions on both the ConvTTS test set and the FDX-Conv OOD set. As shown in Table~\ref{table:pilot-results}, Gemini-2.5-Pro exhibits negative model–human correlation on the ConvTTS test set and only weak correlation on the FDX-Conv OOD set. These findings indicate that, without downstream fine-tuning, frontier speech models still struggle to reliably judge speech naturalness. This observation is consistent with the conclusions of \citet{manakul2025audiojudge}, which identify a notable gap between frontier speech models and human listeners in understanding fine-grained paralinguistic information.

\textbf{Frontier Text Model Acoustic Feature Prompting.}
Motivated by the limitations of frontier speech LLMs in extracting paralinguistic cues, we explore whether frontier \emph{text-only} LLMs can be more effectively leveraged for speech naturalness judgment by conditioning on explicitly extracted acoustic features. For each speech sample, we extract a set of paralinguistic features (details are described in Section~\ref{sec:gsrm}). These features are converted into structured textual descriptions and used to prompt a frontier text LLM to predict sub-metric scores and an human-likeness rating. We evaluate this approach using Gemini-2.5-Pro in text-only mode on both the ConvTTS test set and the FDX-Conv OOD set. As shown in Table~\ref{table:pilot-results}, despite having no direct access to raw audio, the text-based LLM leveraging explicit acoustic feature descriptions consistently outperforms Gemini-2.5-Pro with speech prompting in both in-domain and OOD evaluations. These results suggest that the primary bottleneck in using frontier speech LLMs for speech naturalness judgment lies not in their reasoning capability, but in their ability to reliably extract and attend to fine-grained paralinguistic cues from raw audio. 

\textbf{Remarks.}
These pilot studies suggest that effective speech naturalness prediction requires (i) extracting explicit interpretable paralinguistic cues from waveforms, and (ii) a reasoning mechanism that reliably maps such evidence to perceptual judgments. This motivates the design of the \emph{Generative Speech Reward Model (GSRM)}, which integrates acoustic feature extraction with explicit chain-of-thought reasoning to emulate how humans evaluate speech.
\vspace{-1em}

\section{Generative Speech Reward Model} \label{sec:gsrm}

\begin{figure*}[!t]
    \centering
    \includegraphics[width=1.0\textwidth]{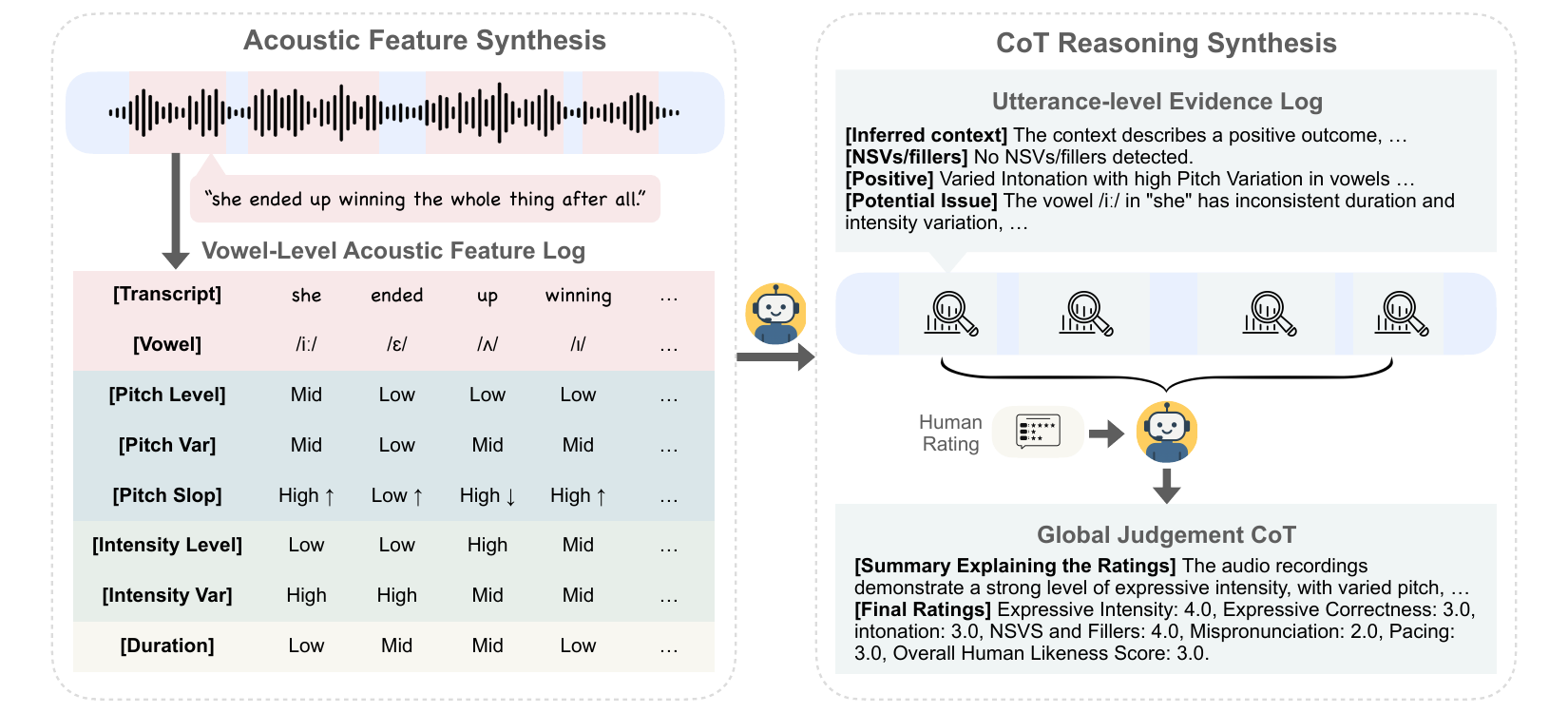}
\caption{\textbf{CoT Synthesis Framework of GSRM.} For each utterance segment, the synthesis pipeline first extracts structured, vowel-level acoustic features from the segment and then synthesizes detailed reasoning to connect paralinguistic cues with perceptual judgments. This process is applied across all utterance segments to construct a complete evidence log for the audio, which is subsequently combined with a global judgment CoT to form the final training trajectory for training GSRM.}
\label{fig:teaser}
\vspace{-0.5em}
\end{figure*}

Our ultimate goal is to build a \emph{generative reward model} tailored for speech, which first extracts inherent paralinguistic evidence from speech and then reasons over this evidence to judge audio quality. While our focus is speech naturalness evaluation, GSRM is designed as a universal framework that can assess any attribute of speech through reasoning. At a high level, GSRM emulates the human evaluation process: human raters implicitly attend to prosodic cues such as pitch movement, loudness variation, and timing, synthesize these observations into quantitative judgments of overall naturalness. To train GSRM to achieve this, we need two stages: (i) extracting structured acoustic features from the raw audio for downstream judgment, and (ii) synthesizing explicit chain-of-thought (CoT) reasoning trajectories that connect these features to quantitative ratings.

\begin{table*}[h]
\centering
\renewcommand{\arraystretch}{1.2}
\setlength{\tabcolsep}{10pt}
\footnotesize
\caption{\textbf{Acoustic Features.}
Vowel-level prosodic features extracted from time-aligned vowel segments, serve as sufficient statistics for speech naturalness judgment.}
\vspace{-0.6em}

\begin{tabular}{p{2.2cm} p{12.6cm}}
\toprule
\textbf{Acoustic Feature} & \textbf{Definition and Interpretation} \\
\midrule
\textbf{Pitch Level} &
Average fundamental frequency ($F_0$) of a vowel segment, reflecting overall pitch height. \\

\textbf{Pitch Var.} &
Within-segment variability of frequency, reflecting the degree of expressiveness. \\

\textbf{Pitch Slope} &
Trend of frequency change across the vowel segment (rising or falling), reflecting local intonation contours. \\

\textbf{Intensity Level} &
Average loudness (e.g., RMS energy) of the vowel segment, reflecting perceived vocal strength or emphasis. \\

\textbf{Intensity Var.} &
Temporal fluctuation of intensity within the vowel segment, indicating stability or variability in loudness. \\

\textbf{Duration} &
Temporal length of the vowel segment, reflecting speech rate, emphasis, and pacing characteristics. \\

\bottomrule
\end{tabular}
\label{table:acoustic-features}
\vspace{-0.5em}
\end{table*}

\begin{table*}[!t]
\centering
\renewcommand{\arraystretch}{1.25}
\setlength{\tabcolsep}{10pt}
\captionsetup{font=small}
\footnotesize
\caption{\textbf{Evidence Log Dimensions.}
Each utterance is analyzed along four dimensions to synthesize interpretable reasoning grounded in acoustic features and transcript context.}
\vspace{-0.6em}

\begin{tabular}{p{2.4cm} p{12.6cm}}
\toprule
\textbf{Dimension} & \textbf{Description} \\
\midrule
\textbf{Inferred Context} &
A concise description of the communicative intent of the utterance (e.g., question, acknowledgment, neutral statement), inferred from the transcript. \\

\textbf{NSVs / Fillers} &
Identification of any non-speech vocalizations or filler words (e.g., ``yeah'', ``uh'',  ``uh''), supported by transcripts and acoustic feature such as abrupt pitch or intensity variation. \\

\textbf{Positive Attributes} &
Identification of perceptual strengths, such as appropriate emphasis, smooth intonation, or natural fillers, grounded in specific vowels and their acoustic features. \\

\textbf{Potential Issue} &
Identification of perceptual weaknesses, including intonation instability, pacing inconsistency, expressive incorrectness, or unnatural pauses, grounded in specific vowels and their acoustic features.  \\

\bottomrule
\end{tabular}
\label{table:cot-dimensions}
\vspace{-1em}
\end{table*}

\textbf{Acoustic Feature Extraction.} Judging speech naturalness directly from raw audio is challenging due to the high dimensionality and redundancy of waveform signals. From an information-theoretic perspective, raw audio contains more information than is required to infer perceptual naturalness ratings. We therefore assume the existence of a set of \emph{sufficient statistics}~\cite{cover1999elements} that preserves all information relevant to naturalness judgment while discarding redundancy. In the context of speech evaluation, such sufficient statistics naturally correspond to a set of interpretable acoustic features. Specifically, we adopt a vowel-level acoustic feature extraction pipeline. Given an audio recording and its transcript, we perform phoneme-level forced alignment to obtain precise temporal boundaries for each phoneme, and retain only vowel segments. This is motivated by the observation that vowels, as voiced and acoustically stable units, carry the majority of prosodic information relevant to expressiveness, intonation, and pacing.

As shown in Figure~\ref{fig:teaser}, for each vowel segment, we extract a compact set of low-level prosodic features as summarized in Table~\ref{table:acoustic-features}. To improve robustness and interpretability, continuous feature values are discretized into ordinal categories (e.g., very low, low, mid, high, very high). After feature extraction, we obtain a raw \emph{acoustic feature log} structured at the utterance level. For each utterance, the log records detailed vowel-level acoustic analyses across the entire utterance, which serve as the main evidence for subsequent reasoning and naturalness judgment.

\textbf{Reasoning Synthesis.} While acoustic features capture essential paralinguistic evidence, they do not, by themselves, specify how such evidence should be interpreted to form perceptual judgments of speech naturalness. Bridging this gap requires the synthesis of CoT reasoning that maps acoustic cues to ratings of each evaluation metrics and overall human-likeness. As shown in Figure~\ref{fig:teaser}, for each utterance in the raw \emph{acoustic feature log}, we provide a frontier text model (GPT-4o) with the utterance transcript and its associated vowel-level acoustic features. The model is prompted to generate a structured analysis along four dimensions: inferred conversational context, identification of non-speech vocalizations or fillers, positive perceptual attributes, and potential issues (see Table~\ref{table:cot-dimensions}). This process produces an \emph{utterance-level evidence log} that articulates how local prosodic patterns contribute to qualitative perceptual impressions. In the second stage, we synthesize a \emph{global judgment CoT} that maps the entire utterance-level evidence log to final human ratings. The model is provided with both the evidence log and the oracle human ratings, and is instructed to generate a coherent assessment explaining how the observed evidence supports the assigned scores. The final CoT trajectory used for training is obtained by concatenating the \emph{utterance-level evidence log} with the synthesized \emph{global judgment CoT}. 

\textbf{Training and Deployment.} We synthesize CoT trajectories for all samples in the ConvTTS training set to construct the training data for GSRM. We then fine-tune Qwen2.5-Omni-7B to take raw audio as input and generate the corresponding synthesized CoT trajectories via supervised fine-tuning (SFT). Through this procedure, the model is encouraged to learn not only \emph{what} ratings to assign, but also \emph{why} those ratings are justified given the underlying acoustic evidence. In addition, we explore whether reinforcement learning with verifiable rewards (RLVR) can further improve naturalness prediction performance. The discussion is provided in Appendix~\ref{sec:gsrm_rl}. Once trained, GSRM is deployed in the same manner as a speech-in, text-out LLMs. Given a raw audio recording, GSRM produces a textual CoT reasoning, followed by numerical ratings for individual sub-metrics as well as an overall human-likeness score. To improve robustness, we further employ test-time scaling by sampling multiple random predictions for the same audio input and averaging the resulting scores. Empirically, we observe that using more test-time samples consistently improves model–human consistency (see Section~\ref{sec:main_results}).

\section{Towards Online RLHF for Speech LLM} \label{sec:rlhf}
\begin{figure}[!t]
    \centering
    \includegraphics[width=0.7\textwidth]{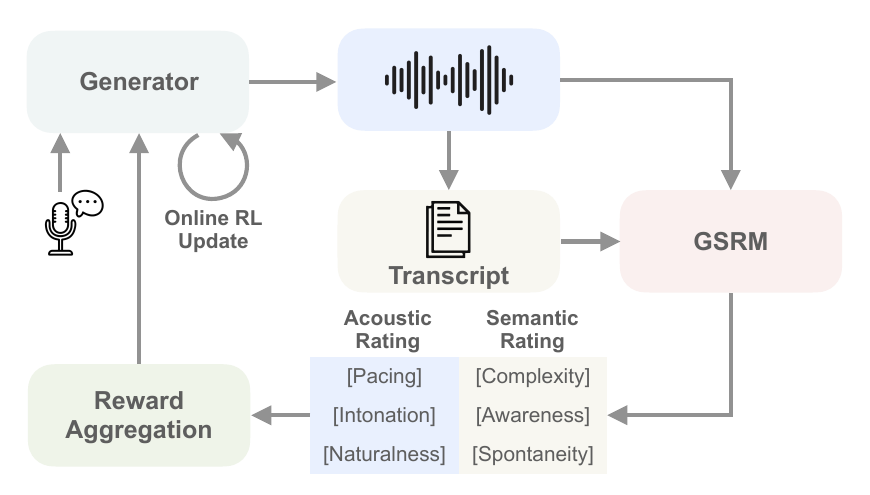}
\caption{\textbf{GSRM for Speech Online RLHF.}
Given a user query, a speech LLM acting as the generator produces a spoken response. Both the generated speech and its text transcript are provided to GSRM, which outputs ratings across multiple acoustic and semantic dimensions. A reward aggregator combines these ratings into a scalar reward that is used as feedback to guide online reinforcement learning updates for the generator.}
\label{fig:rlhf}
\vspace{-1em}
\end{figure}
In this section, we study how to leverage GSRM to improve the speech generation quality of a real-time dialogue system, specifically, an in-house speech LLM through online reinforcement learning from human feedback (RLHF).

\textbf{Extending GSRM to Speech Semantic Judgment.} We first note that speech RLHF poses unique challenges compared to text domain. Spoken responses convey rich information along two intertwined dimensions: \emph{acoustic quality}, including intonation and pacing, and \emph{semantic quality}, such as contextual relevance and coherence. To address this, we extend GSRM beyond acoustic evaluation to also judge the semantic quality of speech. Following the same design principles used for naturalness prediction, we define a set of sub-metrics for speech semantic quality, including language complexity, contextual awareness, and spontaneity. We then collect a diverse training set consisting of textual transcripts of real audio responses generated by our in-house speech LLM. These transcripts are annotated by a multi-agent framework, and high-quality CoT reasoning is synthesized using a teacher model. Finally, we fine-tune GSRM on this synthesized dataset to perform semantic judgment. Additional details are provided in Appendix~\ref{sec:semantic_judge}.

\textbf{Applying GSRM to Online RLHF.} Online reinforcement learning has been widely adopted to improve the reasoning capabilities of text LLMs~\cite{zha2025rl,shen2025satori,guo2025deepseek}, yet its potential for improving the generation quality of speech LLMs remains largely unexplored. A key difference lies in the nature of verification: while RL for reasoning tasks often relies on objective signals, RL for speech generation requires \emph{subjective} evaluation across multiple dimensions. GSRM provides a practical solution to this challenge by serving as a universal verifier. As illustrated in Figure~\ref{fig:rlhf}, given a user query, the speech LLM produces spoken responses as rollouts in a real-time interaction loop. Each speech response is transcribed into text using a lightweight automatic speech recognition (ASR) model, such as Whisper~\cite{radford2023robust}, to disentangle semantic content from acoustic characteristics. The generated speech and its corresponding transcript are then provided to GSRM, which predicts ratings across multiple acoustic and semantic dimensions. These ratings are aggregated into a scalar reward according to a predefined aggregation rule. The generator is then iteratively updated by alternating between producing new speech samples and receiving verification feedback from GSRM via an online RL algorithm.

\section{Experiments} 
\begin{table*}[!t]
\centering
\footnotesize
\captionsetup{font=small}
\caption{\textbf{Main Results.} Correlation metrics are higher-is-better ($\uparrow$); MSE is lower-is-better ($\downarrow$).}
\vspace{-0.5em}
\setlength{\tabcolsep}{5.8pt}
\begin{tabular}{l c
                S[table-format=1.3] S[table-format=1.3] S[table-format=1.3]
                S[table-format=1.3] S[table-format=1.3] S[table-format=1.3]}
\toprule[1.5pt]
& & \multicolumn{3}{c}{\textbf{Validation}} & \multicolumn{3}{c}{\textbf{Out-of-domain}} \\
\cmidrule(lr){3-5}\cmidrule(lr){6-8}
\textbf{Method}
& \textbf{Input}
& {\textbf{Pearson} $\uparrow$}
& {\textbf{Spearman} $\uparrow$}
& {\textbf{MSE} $\downarrow$}
& {\textbf{Pearson} $\uparrow$}
& {\textbf{Spearman} $\uparrow$}
& {\textbf{MSE} $\downarrow$} \\
\midrule
Human Inter-Rater Consistency (Oracle)
& Speech
& 0.533 & 0.517 & 0.301
& 0.532 & 0.489 & 0.336 \\

\midrule
NISQA (Min MOS)
& Speech
& -0.104 & -0.109 & 3.325
& 0.229 & 0.212 & 1.066 \\

NISQA (Mean MOS)
& Speech
& -0.174 & -0.173 & 1.079
& 0.164 & 0.147 & 1.197 \\

UTMOSv2 (Min MOS)
& Speech
& -0.161 & -0.165 & 6.19
& 0.245 & 0.244 & 2.815 \\

UTMOSv2 (Mean MOS)
& Speech
& -0.191 & -0.206 & 0.987
& 0.196 & 0.210 & 0.442 \\

Speech Prompting (Gemini)
& Speech
& -0.050 & -0.078 & 1.009
&  0.140 &  0.136 & 0.531 \\

Acoustic Feature Prompting (Gemini)
& Text
& 0.133 & 0.130 & 1.180
& 0.193 & 0.157 & 0.814 \\

Acoustic Feature Prompting (GPT-4o)
& Text
& 0.100 & 0.103 & 0.549
& 0.277 & 0.214 & 0.317 \\

WavLLM-based Regressor 
& Speech
& 0.331 & 0.279 & 0.242
& 0.222 & 0.212 & 0.275 \\

AES-based Regressor 
& Speech
& 0.384 & 0.352 & 0.228
& 0.236 & 0.224 & 0.265 \\ 

\midrule
\textbf{GSRM (ours)}
& \textbf{Speech}
& \textbf{0.401} & \textbf{0.375} & \textbf{0.332}
& \textbf{0.465} & \textbf{0.427} & \textbf{0.230} \\
\bottomrule
\end{tabular}
\label{table:main-results}
\vspace{-1em}
\end{table*}

\subsection{Settings.} \label{sec:exp_setting}
\textbf{Implementation Details.} We adopt Qwen2.5-Omni-7B~\cite{xu2025qwen2} as the base model for GSRM due to its strong speech understanding capabilities and efficient inference. We synthesize SFT data using the ConvTTS training set. For each audio sample, we average the ratings across annotators and round the result to the nearest integer to obtain a single target score. For reasoning synthesis, we employ GPT-4o~\cite{hurst2024gpt} due to its strong text reasoning and instruction-following capabilities. This process yields 4,579 training examples, each consisting of a raw audio input paired with a synthesized CoT trajectory. We fine-tune the model with full-parameter training for 10 epochs. Additional details are provided in Appendix~\ref{sec:app-setting}.

\textbf{Benchmark and Evaluation.} We evaluate GSRM on both the in-domain ConvTTS test set and the OOD FDX-Conv dataset. Since the ultimate goal of GSRM is to serve as a proxy for human judgment, we focus on measuring consistency between model predictions and human ratings. For each test sample, GSRM generates multiple predictions via random sampling. The final model prediction is obtained by averaging these outputs, which is then compared against the averaged human rating. We report three evaluation metrics: (1) Pearson correlation coefficient (PCC), which measures linear agreement between model predictions and human ratings; (2) Spearman correlation, which captures rank-order consistency and is robust to nonlinear relationships; and (3) mean squared error (MSE), which measures the squared deviation between predictions and human ratings. Unless otherwise stated, all main results are reported using average rating over 16 inference samples per test instance.

\textbf{Baseline Methods.}
In addition to the baselines introduced in Section~\ref{sec:pilot_studies}, we compare GSRM against several established naturalness predictors: (1) NISQA~\cite{mittag2021nisqa}, a CNN-based model for predicting speech naturalness and related perceptual attributes such as noisiness and loudness; (2) UTMOSv2~\cite{baba2024utmosv2}, which combines self-supervised speech representations with spectrogram features to estimate naturalness; and (3) a WavLLM-based regressor, which is fine-tuned on our training data using the WavLLM~\cite{hu2024wavllm}, a self-supervised speech model pre-trained on 94k hours of audio. (4) an AES-based regressor, also fine-tuned on our training data, using the AES Transformer~\cite{tjandra2025meta}, a state-of-the-art model for audio aesthetics assessment.

\subsection{Main Results} \label{sec:main_results}
\textbf{Baseline Comparison.}
The main results are summarized in Table~\ref{table:main-results}. We first observe that existing naturalness predictors struggle to reliably assess speech naturalness, with many methods exhibiting near-zero or even negative model–human correlation on the evaluation datasets. Also, acoustic feature prompting with GPT-4o consistently outperforms speech prompting using Gemini-2.5-Pro, highlighting the effectiveness of explicitly extracted acoustic features for naturalness judgment. While the AES-based predictor achieves competitive performance, its generalization ability on the OOD dataset remains limited. In contrast, GSRM achieves a PCC of $0.465$ on the FDX-Conv OOD set, approaching human inter-rater consistency and substantially outperforming all baseline methods.

\textbf{Scaling Behavior of GSRM.}
\begin{figure}[!t]
    \centering
    \begin{tabular}{rl}
    \subfloat[Scaling Training Data Size]{%
        \includegraphics[width=0.5\textwidth]{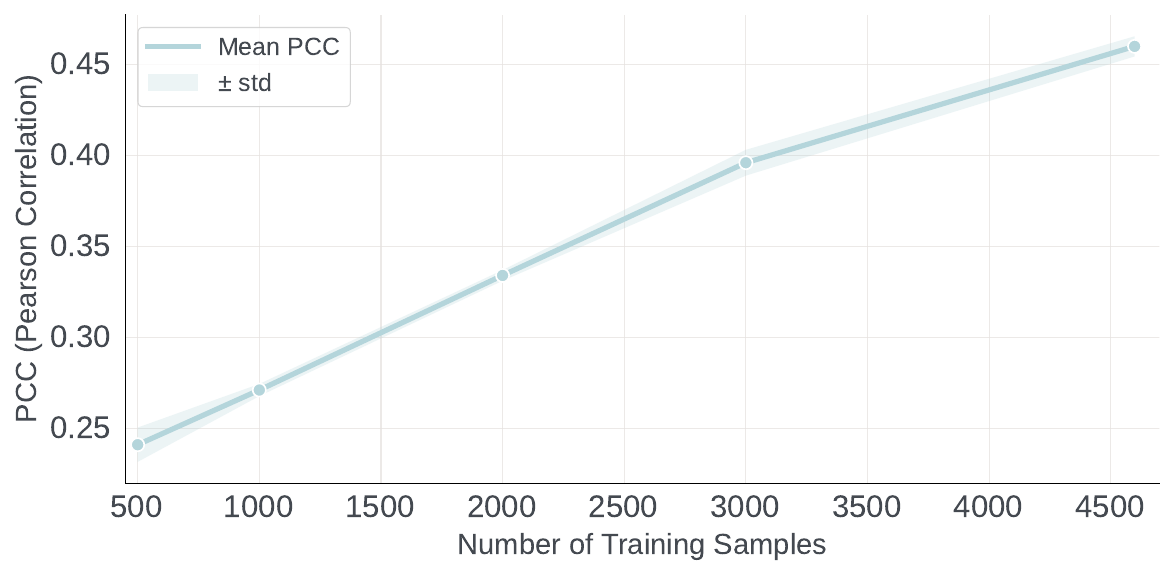}
    } &
    \subfloat[Scaling Test-time Compute]{%
        \includegraphics[width=0.5\textwidth]{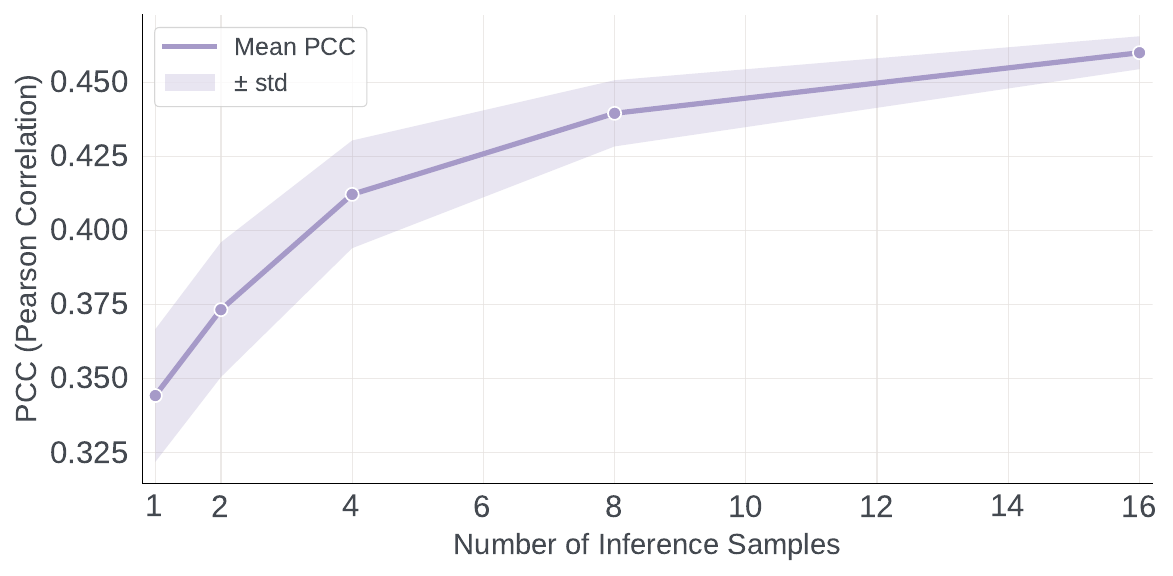}
    }
    \end{tabular}
\caption{\textbf{Scaling Behavior of GSRM.} The y-axis reports PCC performance on the FDX-Conv OOD set.}
\label{fig:scaling_behavior}
\vspace{-1em}
\end{figure}
We further analyze the scaling behavior of GSRM along two dimensions: training data scale and test-time computation. First, we train GSRM with increasing amounts of data ($\{500, 1000, 2000, 3000, 4579\}$ samples) and evaluate its PCC on the FDX-Conv OOD set. As shown in Figure~\ref{fig:scaling_behavior}(a), performance improves steadily with larger training sets, indicating that GSRM can further benefit from more diverse and large-scale human-annotated data. Next, using the model trained on the full dataset, we vary the number of test-time inference samples $K \in \{1, 2, 4, 8, 16\}$ and average the resulting predictions. Figure~\ref{fig:scaling_behavior}(b) shows that PCC increases monotonically as $K$ grows, demonstrating that test-time scaling effectively reduces prediction variance. The most significant gains occur in the low budget regime ($K=1$ to $K=4$), suggesting that robust naturalness prediction can be achieved with a reasonable inference cost.

\subsection{Analysis} \label{sec:analysis}
In this section, we conduct ablation studies to examine the contribution of evidence log, individual acoustic features and evaluation sub-metrics to naturalness prediction.

\textbf{Ablation of Evidence Log.}
\begin{figure}[!t]
    \centering
    \includegraphics[width=0.75\textwidth]{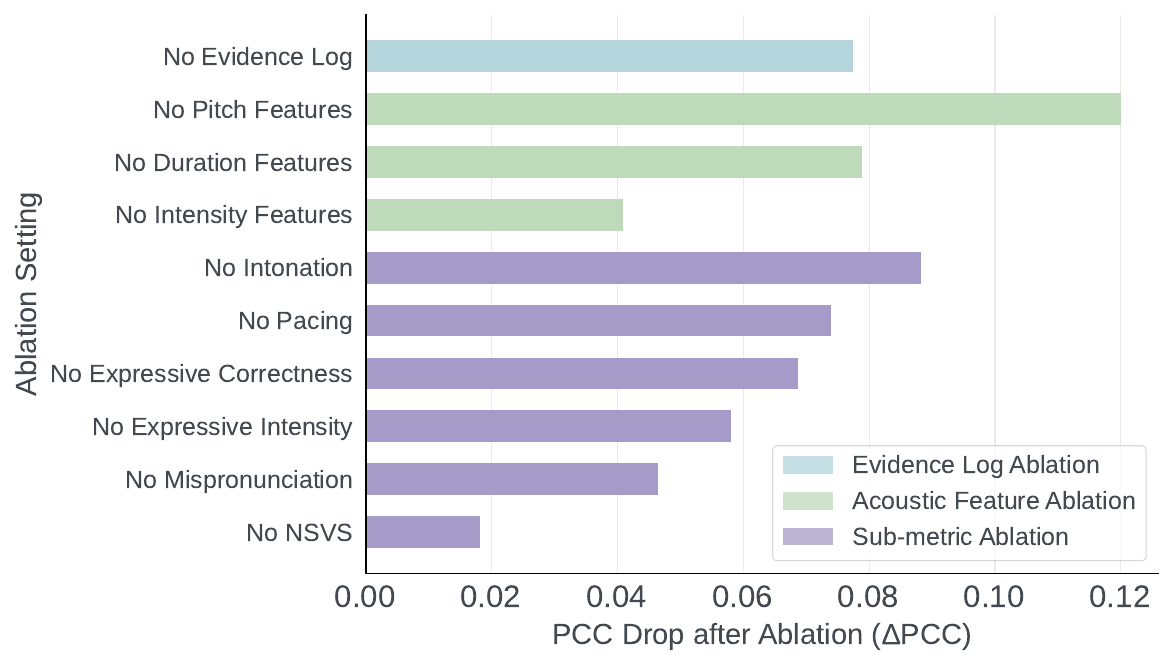}
\caption{\textbf{Ablation Studies of Impact of Sub-metrics and Acoustic Features to Naturalness Prediction.} }
\label{fig:ablation_feature_submetric}
\vspace{-1em}
\end{figure}
As illustrated in Figure~\ref{fig:teaser}, GSRM’s CoT trajectories consist of both an utterance-level evidence log and a global judgment over the entire audio. To isolate the contribution of the evidence log, we ablate it and train GSRM using only the global judgment CoT. As shown in Figure~\ref{fig:ablation_feature_submetric}, removing the evidence log results in a PCC drop of approximately $0.08$. This degradation aligns with our intuition: the utterance-level evidence log captures fine-grained local prosodic patterns, which provide essential context for judging overall speech naturalness.

\textbf{Ablation of Acoustic Feature.}
While evidence log is crucial, acoustic features provide the foundation for generating such evidence log. To assess the importance of different feature groups, we independently ablate pitch-, intensity-, and duration-related features during CoT synthesis and retrain GSRM under each condition. The resulting models are evaluated on the FDX-Conv OOD set. Figure~\ref{fig:ablation_feature_submetric} shows that pitch-based features are the most critical for naturalness prediction, with PCC dropping by more than $0.12$ when pitch information is removed. Duration features exhibit a moderate impact, while intensity-based features contribute comparatively less. These findings align with prior observations that pitch dynamics are a dominant cue in human perception of speech naturalness.

\textbf{Ablation of Sub-metric.}
We next analyze the importance of individual sub-metrics. We keep the acoustic feature log fixed but ablate each sub-metric during CoT synthesis. For example, when the \emph{intonation} sub-metric is removed, GSRM is no longer allowed to reason about intonation when inferring the overall human-likeness score. Figure~\ref{fig:ablation_feature_submetric} indicate that intonation and pacing are the most influential sub-metrics for predicting naturalness, while non-speech vocalizations and fillers (NSVS) contribute the least. These trends are consistent with the correlation analysis of human ratings in Figure~\ref{fig:submetric_pcc}, where intonation and pacing exhibit the highest correlation with human-likeness, and NSVS and mispronunciation show the weakest correlations.

\subsection{Results in the Wild: GSRM for Online RLHF}
\begin{table}[!t]
\centering
\footnotesize
\captionsetup{font=small}
\caption{\textbf{Human A/B Evaluation Results for Online RLHF.} Pairwise preference outcomes comparing the RLHF-trained model against the base model across speech quality dimensions.}
\vspace{-0.5em}
\setlength{\tabcolsep}{3.3pt}
\begin{tabular}{l c c c c}
\toprule[1.5pt]
\textbf{Outcome} 
& \textbf{Tone} 
& \textbf{Pacing} 
& \textbf{Intonation} 
& \textbf{Naturalness} \\
\midrule
Ties 
& 8\%  & 30\% & 18\%  & 6\%  \\
Base Model Wins 
& 18\%  & 10\%  & 16\%  & 12\%  \\
\textbf{RLHF Model Wins} 
& \textbf{74\%} & \textbf{60\%} & \textbf{66\%} & \textbf{82\%} \\
\bottomrule
\end{tabular}
\label{table:rlhf-results}
\vspace{-1em}
\end{table}
Finally, we evaluate the effectiveness of GSRM in improving the naturalness of speech generation in an in-house speech LLM through online RLHF. We start from an in-house speech LLM trained with supervised fine-tuning (SFT), which serves as the base model. We then curate an RL training set consisting of 9.2k speech prompts. For each prompt, the model is asked to generate a single-turn speech response, which is subsequently evaluated by GSRM along multiple sub-dimensions. Additional details of the online RLHF training are provided in Appendix~\ref{sec:rlhf_details}. After training, we compare the RLHF-trained model with the base SFT model. Specifically, both models generate speech responses for a held-out test prompt set, and human raters conduct A/B evaluations across multiple dimensions, including tone, pacing, intonation, and overall naturalness. Each speech sample is evaluated by five independent human raters to ensure robustness. Table~\ref{table:rlhf-results} demonstrate that the RLHF-trained model outperforms the base model in terms of overall naturalness, winning the A/B comparisons on 82\% of the test samples. Moreover, RLHF training with GSRM consistently improves speech generation quality across other sub-dimensions, further highlighting the effectiveness of GSRM as a reward signal for online RLHF.
\section{Concluding Remarks}
We introduced the \emph{Generative Speech Reward Model}, a reasoning-centric reward model tailored for speech naturalness evaluation and reinforcement learning from human feedback (RLHF). Our empirical results demonstrate that GSRM substantially outperforms existing speech naturalness predictors. Importantly, we also show that GSRM can serve as an effective verifier to improve the generation quality of a speech language model. Looking ahead, we view GSRM as a foundational component for future speech alignment systems. Beyond naturalness, the framework can be naturally extended to evaluate and optimize additional speech attributes, such as emotional expressivity, style consistency, and speaker similarity. More broadly, a promising future direction is to explore advanced speech RLHF paradigms that jointly co-evolve the GSRM and the generator within a unified RL framework, as has proven successful in reasoning domains~\cite{zha2025rl}.

\section*{Acknowledgment}
The authors are listed according to their respective contribution roles.

\textbf{Core Contributors.}
Among the four core contributors, Maohao and Naoyuki jointly initiated the high-level idea of training GSRM for speech RLHF. Osama proposed the acoustic feature extraction method, and Maohao suggested that acoustic features are essential for GSRM. Tejas extended the GSRM framework to the semantic judge setting and conducted the human evaluation. All core contributors were directly involved in the implementation of both the GSRM training pipeline and the speech LLM RLHF pipeline.

\textbf{Contributors.}
Yancheng contributed to the acoustic feature extraction pipeline. Kateřina, Ruiming, and Niko provided support for the RL training infrastructure. Anfeng and Yashesh contributed to the development of the regressor-based baseline methods.

\textbf{Project Leads.}
Gregory, Qing, and Jilong served as the senior supervisors of this project. Jilong served as Maohao’s internship manager, providing the necessary resources and overseeing the entire project.

\textbf{Data Support.}
The authors would like to thank Wei-Ning Hsu and Ann Lee for their support with the dataset.


\clearpage
\newpage
\bibliographystyle{assets/plainnat}
\bibliography{paper}

\clearpage
\newpage
\beginappendix
\begingroup
\hypersetup{linkcolor=darkblue} 
\addtocontents{toc}{\protect\StartAppendixEntries}
\listofatoc
\endgroup

\section{Related Work} \label{sec:related}

\subsection{Speech and Audio Quality Prediction.}
Automatic speech naturalness assessment has been widely studied through non-intrusive MOS predictors that regress perceptual scores directly from audio. Early neural models such as MOSNet~\cite{lo2019mosnet} demonstrated the feasibility of predicting naturalness without reference signals, while subsequent work improved robustness and coverage across speech synthesis and telephony scenarios, including NISQA~\cite{mittag2021nisqa} and DNSMOS~\cite{reddy2021dnsmos}. More recent approaches, such as UTMOSv2~\cite{baba2024utmosv2}, leverage large-scale self-supervised speech representations to better match human judgments across domains. Beyond speech-specific MOS prediction, audio aesthetics and quality models aim to assess broader perceptual attributes across speech, music, and sound, such as Meta-AES~\cite{tjandra2025meta} and the AudioMOS benchmark~\cite{huang2025audiomos}. Despite strong empirical performance, these models operate as black-box regressors that output scalar scores without interpretable justification, motivating the need for generative reward models that explicitly reason over paralinguistic evidence.

\subsection{Generative Reward Models.}
Reward modeling has recently shifted from discriminative scoring toward \emph{generative} formulations that produce explicit critiques or reasoning before assigning rewards. This paradigm includes LLM-as-a-judge approaches~\cite{li2025generation,li2024llms} and generative verifiers that cast reward modeling as a reasoning task~\cite{zhang2024generativeverifiers}. DeepSeek further introduces inference-time scaling for generalist generative reward models, demonstrating that aggregating multiple independently generated judgments significantly improves robustness and generalization~\cite{liu2025inference}. Related work such as GenRM~\cite{mahan2024generative} and J1~\cite{whitehouse2025j1} shows that generative reward models can be iteratively improved via self-generated rationales and online reinforcement learning. While these approaches are now well-established in text domains, analogous techniques for speech and audio remain under-explored.

In the speech domain, AudioJudge~\cite{manakul2025audiojudge} evaluates the capability of frontier speech LLMs to judge audio quality and paralinguistic attributes through prompt engineering, revealing that most models struggle to provide reliable judgments. The most closely related concurrent works to GSRM are SpeechJudge~\cite{zhang2025speechjudge} and WavReward~\cite{ji2025wavreward}, which train speech LLMs to generate chain-of-thought rationales for pairwise speech comparisons. While effective, both approaches rely on a teacher speech LLM to generate reasoning, which can inherit whatever the teacher model has encoded about naturalness, despite its reliability. In contrast, GSRM decomposes reward modeling into explicit acoustic feature extraction followed by feature-grounded generative reasoning, aligning more closely with how humans assess fine-grained paralinguistic cues.

\subsection{RLHF for Speech and Audio.}
Applying reinforcement learning from human feedback (RLHF) to speech and audio generation is challenging due to the continuous and high-dimensional nature of audio signals, yet recent work has demonstrated promising progress, primarily in text-to-speech (TTS) settings. SpeechJudge~\cite{manakul2025audiojudge} applies direct preference optimization (DPO)~\cite{rafailov2023direct} to post-train TTS models using human preferences. Other studies improve TTS quality using learned or proxy reward signals, including diffusion-based reinforcement learning such as DLPO~\cite{chen2024dlpo} and offline preference optimization for text-to-audio alignment~\cite{liao2024baton}. Additionally, ASR-based rewards have been explored to enhance intelligibility and perceived naturalness of speech model outputs~\cite{liu2025group}. Despite these advances, existing RL-based approaches focus almost exclusively on single-utterance TTS or offline audio generation, and no prior work has investigated leveraging \emph{online} reinforcement learning to improve the naturalness of full-duplex, interactive speech LLMs.

\section{GSRM Demo Examples} \label{sec:demo}
We present demo examples to illustrate how GSRM analyzes spoken responses and produces interpretable judgments of human-likeness. Given any audio clip, the model generates an evidence log that explicitly grounds its judgments in acoustic cues such as pitch variation, pitch slope, intensity, and duration, and aggregates these signals into sub-metric ratings and an overall score. Across examples with increasing ground-truth human-likeness ratings, the model exhibits consistent behavior: lower-rated samples tend to show unstable expressiveness, exaggerated pitch variation, or uneven pacing, while higher-rated samples demonstrate more coherent intonation, appropriate expressive intensity, and smoother temporal dynamics. 

\begin{tcolorbox}[blue_box, title = {{Audio Sample 1 : Ground-truth Human-likeness Rating = 2.0}}]\footnotesize
[Evidence log]

\#\# Transcript: ``oh, that's so sweet."

[Inferred context]  

The context inferred is celebration, as the phrase ``oh, that's so sweet" suggests a positive reaction to something delightful or heartwarming.

[NSVs/fillers]  

No NSVs/fillers detected.

[Positive]  

The audio exhibits expressive intensity with high pitch variation and high falling pitch slopes, particularly in the vowels /\IPA{o}/ in ``oh" and /\IPA{iː}/ in ``sweet," indicating a dynamic and engaging delivery.

[Potential Issue]  

No notable issues detected.

\#\# Transcript: ``i think i just had a tiny bit of fun imagining that puppy's adventures."

[Inferred context]  

The context inferred is a neutral statement, as the transcript describes a personal experience of imagining a puppy's adventures without any emotional cues.

[NSVs/fillers]  

No NSVs/fillers detected.

[Positive]  

The audio exhibits varied intonation with high pitch variation and clear pitch slopes, particularly in the vowels /\IPA{a}/ in ``i" and /\IPA{æ}/ in ``had," indicating expressive intensity.

[Potential Issue]  

There are potential intonation problems due to the extremely high pitch in the vowel /\IPA{ɪ}/ in ``think," which may lead to unstable expressiveness.

\#\# Transcript: ``did i get it right?"

[Inferred context]  

The context inferred is a question, as indicated by the transcript ``did i get it right?" which typically seeks confirmation or clarification.

[NSVs/fillers]  

No NSVs/fillers detected.

[Positive]  

The audio demonstrates good expressive intensity with clear pitch slopes and moderate pitch variation across the vowels, contributing to a natural and engaging delivery.

[Potential Issue]  

No notable issues detected.

\#\# Transcript: ``i'm all ears."

[Inferred context]  

The phrase ``I'm all ears" suggests an empathetic context, indicating the speaker is ready to listen attentively.

[NSVs/fillers]  

No NSVs/fillers detected.

[Positive]  

The vowel /\IPA{iː}/ in ``ears" shows varied intonation with high pitch variation and a high rising pitch slope, contributing to expressive intensity.

[Potential Issue]  

The vowel /\IPA{ɔː}/ in ``all" has very low pitch variation and intensity variation, which may lead to flat intonation and reduced expressiveness.

\#\# Transcript: ``so the puppy was found by a kind family."

[Inferred context]  

The context inferred is a neutral statement, as the transcript describes a factual event about a puppy being found by a kind family.

[NSVs/fillers]  

No NSVs/fillers detected.

[Positive]  

The audio exhibits good expressive intensity with high pitch variation and clear pitch slopes in vowels like /\IPA{o}/ in ``so" and /\IPA{ʌ}/ in ``was," contributing to a dynamic delivery.

[Potential Issue]  

The vowel /\IPA{iː}/ in ``family" shows very low intensity and low rising pitch slope, which may indicate mispronunciation or weak articulation.

\#\# Transcript: ``they gave him a warm bath, a cozy bed, and lots of love."

[Inferred context]  

The context inferred is a celebration or expression of care, as the transcript describes giving someone a warm bath, a cozy bed, and lots of love.

[NSVs/fillers]  

No NSVs/fillers detected.

[Positive]  

The audio demonstrates good expressive intensity, with varied pitch slopes and moderate to high intensity in vowels like /\IPA{e}/ in ``gave" and /\IPA{ɒ}/ in ``lots," contributing to a lively and engaging delivery.

[Potential Issue]  

There are potential pacing issues, as the very high duration of vowels like /\IPA{æ}/ in ``bath" and /\IPA{ɛ}/ in ``bed" may suggest uneven pacing.

\#\# Transcript: ``he settled in quickly and became a beloved member of the family."

[Inferred context]

The context inferred is a neutral statement, as the transcript describes a factual situation about someone becoming a beloved member of a family.

[NSVs/fillers]

No NSVs/fillers detected.

[Positive]

The audio exhibits good expressive intensity with consistent high falling pitch slopes across most vowels, such as /\IPA{ə}/ in ``a" and /\IPA{ʌ}/ in ``of," indicating a clear and expressive delivery.

[Potential Issue]

There are potential pacing issues, as evidenced by the very high duration of the vowel /\IPA{ʌ}/ in ``love" compared to other vowels, which may suggest uneven pacing. \\

[Summary Explaining the Ratings]

The audio recordings exhibit a range of expressive qualities, with some strengths in expressiveness and intonation, but also some areas that could be improved. The expressiveness is generally dynamic, with varied pitch and intensity, although there are instances where expressiveness could be more stable. Intonation is mostly clear, but there are occasional issues with pitch variation that affect the naturalness of the delivery. The pacing is generally natural, but there are moments where it becomes uneven. No non-speech vocalizations or fillers were detected, and there are no clear mispronunciations, although some articulation could be stronger.

expressive\_intensity: The audio demonstrates a good level of expressive intensity, with high pitch variation and clear pitch slopes in several instances, such as in ``oh, that's so sweet" and ``they gave him a warm bath." However, there are moments where expressiveness could be more consistent, such as in ``i think i just had a tiny bit of fun imagining that puppy's adventures," where the pitch in ``think" is extremely high. Therefore, a rating of 3 is appropriate.

expressive\_correctness: The expressiveness is generally appropriate and correct for the context, with the delivery matching the inferred emotional tone of the statements. For example, the phrase ``oh, that's so sweet" is delivered with a positive and engaging tone. However, the instability in expressiveness in some parts, like the high pitch in ``think," slightly detracts from the overall correctness. Thus, a rating of 4 is suitable.

intonation: The intonation is mostly clear and contributes to a natural delivery, with good pitch variation in many parts. However, there are some issues, such as the flat intonation in ``i'm all ears" due to low pitch variation in ``all." These minor issues suggest a rating of 4.

nsvs\_and\_fillers: No non-speech vocalizations or fillers were detected throughout the audio, indicating a clean and natural delivery in this aspect. Therefore, a rating of 5 is warranted.

mispronunciation: There are no clear mispronunciations detected, although some articulation could be stronger, such as in ``family." However, this does not constitute a mispronunciation. Thus, a rating of 2 is appropriate.

pacing: The pacing is generally natural, but there are moments of uneven pacing, such as the high duration of vowels in ``bath" and ``bed," which suggest potential pacing issues. Therefore, a rating of 3 is fitting.

overall\_score: Taking into account all factors, including expressiveness, intonation, and pacing, the overall human likeness of the audio is somewhat limited by the inconsistencies in expressiveness and pacing. Therefore, an overall score of 2 is appropriate. \\

[Final Ratings]

expressive\_intensity: 3.0  

expressive\_correctness: 4.0  

intonation: 4.0  

nsvs\_and\_fillers: 5.0  

mispronunciation: 2  

pacing: 3.0  

overall\_score: 2.0
\end{tcolorbox}

\begin{tcolorbox}[blue_box, title = {{Audio Sample 2 : Ground-truth Human-likeness Rating = 3.0}}]\footnotesize
[Evidence log]

\#\# Transcript: ``the capital of france is paris."

[Inferred context]  

No clear context inferred.

[NSVs/fillers]  

No NSVs/fillers detected.

[Positive]  

The vowel /\IPA{æ}/ in ``France" shows varied intonation with very high pitch variation and a high falling pitch slope, indicating expressive intensity.

[Potential Issue]  

The vowel /\IPA{ɪ}/ in ``is" has very low intensity and very high intensity variation, which may suggest pacing issues or unstable expressiveness.

\#\# Transcript: ``the capital of germany is berlin."

[Inferred context]  

No clear context inferred.

[NSVs/fillers]  

No NSVs/fillers detected.

[Positive]  

The audio demonstrates varied intonation with high falling pitch slopes across vowels, contributing to a dynamic delivery.

[Potential Issue]  

The low pitch and high pitch variation in the vowel /\IPA{ɪ}/ in ``is" may indicate unstable expressiveness, and the very low intensity variation in the vowel /\IPA{ɛ}/ in ``Germany" suggests a lack of emphasis.

\#\# Transcript: ``sounds like fun."

[Inferred context]  

The context inferred is a celebration, as the phrase ``sounds like fun" suggests excitement or enjoyment.

[NSVs/fillers]  

No NSVs/fillers detected.

[Positive]  

The audio exhibits expressive intensity with high pitch variation and clear pitch slopes, particularly in the vowels /\IPA{a}/ and /\IPA{ʌ}/, indicating a lively and engaging delivery.

[Potential Issue]  

The very high pitch variation and intensity in the vowel /\IPA{ʌ}/ in ``fun" may suggest slightly exaggerated expressiveness, potentially leading to unstable intonation.

\#\# Transcript: ``i'll go when you're ready."

[Inferred context]  

The context inferred is instruction, as the sentence ``i'll go when you're ready" suggests a commitment to wait for someone's readiness.

[NSVs/fillers]  

No NSVs/fillers detected.

[Positive]  

The audio exhibits varied intonation with high falling pitch slopes in several vowels, such as /\IPA{a}/ in ``I'll" and /\IPA{o}/ in ``go," indicating expressive intensity.

[Potential Issue]  

There is a potential issue with expressive correctness due to the very high intensity variation in the vowel /\IPA{a}/ in ``I'll," which may suggest unstable expressiveness.

\#\# Transcript: ``three."

[Inferred context]  

No clear context inferred.

[NSVs/fillers]  

No NSVs/fillers detected.

[Positive]  

The high falling pitch slope and moderate pitch variation in both vowels suggest good expressive intensity and intonation.

[Potential Issue]  

The very high duration of the vowel /\IPA{iː}/ in ``three" may indicate slow pacing.

\#\# Transcript: ``got it."

[Inferred context]  

No clear context inferred.

[NSVs/fillers]  

No NSVs/fillers detected.

[Positive]  

The high Pitch Variation and clear high falling Pitch Slope of the vowel /\IPA{ɒ}/ in ``got" suggest expressive intensity and good emphasis.

[Potential Issue]  

The very high Pitch Variation may indicate unstable or exaggerated expressiveness, potentially affecting expressive correctness.

\#\# Transcript: ``let's start fresh."

[Inferred context]  

The context inferred is instruction, as the phrase ``let's start fresh" suggests initiating a new beginning or process.

[NSVs/fillers]  

No NSVs/fillers detected.

[Positive]  

The audio exhibits expressive intensity with high pitch variation and clear pitch slopes, particularly in the vowels /\IPA{ɛ}/ and /\IPA{ɑː}/, indicating a dynamic delivery.

[Potential Issue]  

The vowel /\IPA{ɛ}/ in ``fresh" shows low pitch and mid intensity, which may lead to a slightly flat intonation compared to the rest of the sentence.

\#\# Transcript: ``what's the first number?"

[Inferred context]  

The context inferred is a question, as indicated by the transcript ``what's the first number?" which typically seeks information.

[NSVs/fillers]  

No NSVs/fillers detected.

[Positive]  

The audio demonstrates good expressive intensity with moderate to high pitch variation and clear pitch slopes, particularly in the vowels /\IPA{ʌ}/ in ``what's" and /\IPA{ɜ˞}/ in ``first."

[Potential Issue]  

The vowel /\IPA{ə}/ in ``the" shows a high falling pitch slope with low intensity, which may suggest a lack of emphasis or potential mispronunciation.

\#\# Transcript: ``three, six, 11, 21."

[Inferred context]  

No clear context inferred.

[NSVs/fillers]  

No NSVs/fillers detected.

[Positive]  

The vowel /\IPA{ɪ}/ in ``six" shows varied intonation with high pitch variation and a high falling pitch slope, indicating expressive intensity.

[Potential Issue]  

The vowel /\IPA{iː}/ in ``three" has very low intensity and very high intensity variation, suggesting potential mispronunciation or unstable expressiveness. \\

[Summary Explaining the Ratings]

The audio recordings exhibit a range of expressive intensity, with some instances of high pitch variation and clear pitch slopes contributing to a dynamic delivery. However, there are also moments of unstable expressiveness and potential pacing issues, particularly in the vowels of certain words. The intonation is generally varied, but there are instances where it may lead to instability or flatness. The absence of non-speech vocalizations and fillers is a positive aspect, contributing to the naturalness of the speech. Mispronunciation is present, as indicated by low intensity and potential issues in certain vowels. The pacing is mostly natural, though there are moments of slow pacing. Overall, the audio demonstrates a moderate level of human likeness, with room for improvement in stability and pronunciation.

expressive\_intensity: The audio shows varied expressive intensity, with high pitch variation and clear pitch slopes in several instances, such as in ``France" and ``fun." However, there are also moments of exaggerated expressiveness, leading to a moderate rating. Rating: 3.0

expressive\_correctness: While the expressiveness is generally appropriate, there are instances of unstable expressiveness, such as in ``is" and ``I'll," which affect the correctness. The expressiveness is mostly suitable for the context, but there are occasional issues. Rating: 3.0

intonation: The intonation is varied, with high pitch variation contributing to a dynamic delivery. However, there are moments of instability, such as in ``got" and ``fresh," which affect the overall quality. Rating: 3.0

nsvs\_and\_fillers: The absence of non-speech vocalizations and fillers is a positive aspect, contributing to the naturalness of the speech. Rating: 4.0

mispronunciation: There are instances of potential mispronunciation, such as in ``the" and ``three," which affect the clarity of the speech. Rating: 1.0

pacing: The pacing is mostly natural, but there are moments of slow pacing, such as in ``three," which affect the overall flow. Rating: 3.0

overall\_score: Taking into account the varied expressiveness, intonation, absence of fillers, presence of mispronunciation, and pacing issues, the overall human likeness of the audio is moderate. Rating: 3.0 \\

[Final Ratings]

expressive\_intensity: 3.0  

expressive\_correctness: 3.0  

intonation: 3.0  

nsvs\_and\_fillers: 4.0  

mispronunciation: 1.0  

pacing: 3.0  

overall\_score: 3.0

\end{tcolorbox}

\begin{tcolorbox}[blue_box, title = {{Audio Sample 3 : Ground-truth Human-likeness Rating = 4.0}}]\footnotesize
[Evidence log]

\#\# Transcript: ``so there's this woman and she's at a garage sale, right?"

[Inferred context]  

The context appears to be a neutral statement, as the transcript describes a scenario without emotional cues.

[NSVs/fillers]  

No NSVs/fillers detected.

[Positive]  

The audio demonstrates varied intonation with high pitch variation in vowels like /\IPA{ɪ}/ in ``this" and /\IPA{e}/ in ``sale," contributing to expressive intensity.

[Potential Issue]  

There are potential pacing issues, as the vowel /\IPA{æ}/ in ``and" has low intensity and high intensity variation, which may indicate uneven pacing.

\#\# Transcript: ``she finds this old painting that she's really drawn to."

[Inferred context]  

The context inferred is a neutral statement, as the transcript describes an observation about a painting without emotional cues.

[NSVs/fillers]  

No NSVs/fillers detected.

[Positive]  

The audio demonstrates expressive intensity with high pitch variation and clear pitch slopes in vowels like /\IPA{iː}/ in ``she" and /\IPA{a}/ in ``finds," contributing to a dynamic delivery.

[Potential Issue]  

The vowel /\IPA{o}/ in ``old" shows very low pitch variation and low intensity, which may indicate a flat intonation or lack of emphasis in that part of the sentence.

\#\# Transcript: ``she buys it for a few bucks and takes it home."

[Inferred context]  

No clear context inferred.

[NSVs/fillers]  

No NSVs/fillers detected.

[Positive]  

The audio demonstrates varied intonation with high falling pitch slopes in most vowels, contributing to expressive intensity.

[Potential Issue]  

The vowel /\IPA{æ}/ in ``and" shows very high intensity variation, which may indicate unstable expressiveness or pacing issues.

\#\# Transcript: ``as she's cleaning it, she notices something weird."

[Inferred context]  

The context inferred is a neutral statement, as the transcript describes an observation without emotional cues.

[NSVs/fillers]  

No NSVs/fillers detected.

[Positive]  

The audio exhibits varied intonation with high pitch variation and clear pitch slopes in vowels like /\IPA{o}/ in ``notices" and /\IPA{ʌ}/ in ``something," contributing to expressive intensity.

[Potential Issue]  

There are potential pacing issues, as the duration of vowels like /\IPA{iː}/ in ``weird" is long, while others like /\IPA{æ}/ in ``as" are very low, indicating inconsistent pacing.

\#\# Transcript: ``there's this small compartment hidden behind the canvas."

[Inferred context]  

No clear context inferred.

[NSVs/fillers]  

No NSVs/fillers detected.

[Positive]  

The varied intonation, particularly the high rising pitch slope in ``small" and ``compartment," suggests expressive intensity and good emphasis.

[Potential Issue]  

The very low intensity variation in ``there's" and low pitch variation in ``canvas" may indicate flat intonation and lack of expressiveness in parts of the sentence.

\#\# Transcript: ``she opens it and finds a note from the artist."

[Inferred context]  

No clear context inferred.

[NSVs/fillers]  

No NSVs/fillers detected.

[Positive]  

The varied intonation, particularly the high rising pitch in ``opens" and mid rising pitch in ``artist," contributes to a dynamic and engaging delivery.

[Potential Issue]  

The very low intensity in ``and" and low intensity in ``the" may lead to a flat delivery in parts of the sentence, potentially affecting expressiveness.

\#\# Transcript: ``it's a pretty cool story."

[Inferred context]  

The context inferred is a neutral statement, as the transcript ``it's a pretty cool story" suggests a simple sharing of information without emotional cues.

[NSVs/fillers]  

No NSVs/fillers detected.

[Positive]  

The audio demonstrates good expressive intensity with high falling pitch slopes across vowels, particularly in ``it" and ``pretty," indicating a clear and engaging delivery.

[Potential Issue]  

There is a potential issue with expressive correctness, as the very high intensity variation in ``it's" may suggest an exaggerated or inconsistent expressiveness. \\

[Summary Explaining the Ratings]

The audio recordings demonstrate a high level of expressive intensity, with varied intonation and pitch variation contributing to a dynamic delivery. However, there are some issues with pacing and potential mispronunciations that affect the overall quality. The absence of non-speech vocalizations and fillers is a positive aspect, but the presence of mispronunciations detracts from the overall human likeness. The pacing is somewhat inconsistent, with variations in intensity and duration of vowels, which could affect the natural flow of the speech.

expressive\_intensity: The audio consistently shows high expressive intensity, with varied intonation and pitch variation across most sentences. This is evident in the high pitch variation in vowels like /\IPA{ɪ}/ in ``this" and /\IPA{e}/ in ``sale," as well as the high falling pitch slopes in ``opens" and ``artist." These features contribute to a dynamic and engaging delivery. Rating: 5.0

expressive\_correctness: The expressiveness is appropriate and correct for the context, with the varied intonation and pitch slopes aligning well with the neutral statements being made. There is a slight issue with the very high intensity variation in ``it's," which may suggest exaggerated expressiveness, but overall, the expressiveness is suitable. Rating: 5.0

intonation: The intonation quality is strong, with varied pitch slopes and high pitch variation contributing to a clear and engaging delivery. The high rising pitch slope in ``small" and ``compartment" and the high falling pitch slopes in ``it" and ``pretty" are examples of effective intonation. Rating: 5.0

nsvs\_and\_fillers: There are no non-speech vocalizations or fillers detected in the audio, which contributes to a clean and professional delivery. This absence is a positive aspect of the recordings. Rating: 5.0

mispronunciation: There is evidence of mispronunciation in the audio, which affects the clarity and naturalness of the speech. This is a significant issue that detracts from the overall quality. Rating: 1

pacing: The pacing is somewhat inconsistent, with variations in intensity and duration of vowels indicating uneven pacing. For example, the vowel /\IPA{æ}/ in ``and" shows high intensity variation, and the duration of vowels like /\IPA{iː}/ in ``weird" is long, while others like /\IPA{æ}/ in ``as" are very low. These issues affect the natural flow of the speech. Rating: 4.0

overall\_score: Taking into account the high expressive intensity, appropriate expressiveness, strong intonation, absence of fillers, presence of mispronunciations, and inconsistent pacing, the overall human likeness of the audio is good but not perfect. The mispronunciations and pacing issues prevent a higher rating. Rating: 4.0 \\

[Final Ratings]

expressive\_intensity: 5.0

expressive\_correctness: 5.0

intonation: 5.0

nsvs\_and\_fillers: 5.0

mispronunciation: 1

pacing: 4.0

overall\_score: 4.0
\end{tcolorbox}
\section{Experimental Setup} \label{sec:app-setting}

\subsection{Implementation Details of Baseline Methods}

\subsubsection{Direct Score Predictor.}
We fine-tune Qwen2.5-Omni-7B on the ConvTTS training set, which consists of 4,579 audio samples. Given an input speech recording, the model is trained to directly output numerical ratings for all evaluation sub-metrics as well as the overall human-likeness score, without generating any intermediate chain-of-thought (CoT) reasoning. The supervised fine-tuning (SFT) prompt template used for this baseline is shown below.

We use a learning rate of $2\times10^{-5}$ with a cosine learning rate scheduler. The batch size is set to 32, the maximum sequence length is 4096, and the model is trained for up to 10 epochs. All experiments are conducted using the SWIFT framework~\cite{zhao2024swiftascalablelightweightinfrastructure}.
\begin{tcolorbox}[gray_box, title = {{Prompt Template: Direct Score Predictor}}]\footnotesize
You are an expert audio rater specializing in evaluating the quality of audio recordings.

You will be provided with an audio recording of a conversation between a TTS system and a user. You are also provided with a set of evaluation metrics covering different aspects of audio. 

Your task is to write a natural, clear, and concise assessment of each evaluation metric and assign a rating based on your listening experience. \\

\# Evaluation Metrics

Each evaluation metric is described below:

- expressive_intensity: The intensity of expressiveness in the audio.

- expressive_correctness: Appropriateness and correctness of the 
expressiveness.

- intonation: Quality of the intonation in the speech.

- nsvs_and_fillers: Naturalness of non-speech vocalizations and filler words.

- mispronunciation: whether mispronunciation is present.

- pacing: Naturalness of the pacing or speed of the speech.

- overall_score: Overall Human likeness rating

Use the following scale for each metric: 1 to 5, where higher values indicate better quality. Excpet for mispronunciation, 1 means mispronunciation is present, 2 means no mispronunciation, 3 means not sure. \\

\# Instruction and Response Format

Please follow these steps to complete the task. Your response must strictly follow this format:

[Final Ratings]

expressive_intensity: <your rating>

expressive_correctness: <your rating>

intonation: <your rating>

nsvs_and_fillers: <your rating>

mispronunciation: <your rating>

pacing: <your rating>

overall_score: <your rating>

\end{tcolorbox}

\subsubsection{Few-shot Speech Prompting.}
Following the setup of \citet{manakul2025audiojudge}, we randomly select five audio samples with human-likeness ratings spanning 1 to 5 as in-context learning (ICL) examples, ensuring coverage from low to high-quality speech. The corresponding audio clips are concatenated into a single audio sequence, with short segments of background silence inserted between clips. We prompt Gemini-2.5-Pro in voice mode using this stitched audio together with a few-shot prompt that includes the ground-truth human ratings for the five examples. The full prompt template is provided below.
\begin{tcolorbox}[gray_box, title = {{Prompt Template: Few-shot Speech Prompting}}]\footnotesize
You are an expert audio rater specializing in evaluating the quality of audio recordings.

You will be provided with an audio recording of five demonstration conversation segments between a TTS system and a user. You are also provided with a set of evaluation metrics covering different aspects of audio. 

Your task is to write a natural, clear, and concise assessment of each evaluation metric and assign a rating to the final conversation segment (the sixth segment).

Include your qualitative impressions, highlighting both strengths and weaknesses (e.g., drifts in pitch style across turns, or flatness) of the audio. 

Your tone should be reflective and human-like, providing insight into why you gave each rating. Each claim must be supported with evidence from the audio recording. \\

\# Evaluation Metrics

Each evaluation metric is described below:

- expressive_intensity: The intensity of expressiveness in the audio.

- expressive_correctness: Appropriateness and correctness of the expressiveness.

- intonation: Quality of the intonation in the speech.

- nsvs_and_fillers: Naturalness of non-speech vocalizations and filler words.

- mispronunciation: whether mispronunciation is present.

- pacing: Naturalness of the pacing or speed of the speech.

- overall_score: Overall Human likeness rating of the second speaker (speaker-2), take all the above metrics into account.

Use the following scale for each metric: 1-5. Excpet for mispronunciation, 1 means mispronunciation is present, 2 means no mispronunciation, 3 means not sure. \\

\# Response Format

Please follow these steps to complete the task. Your response must strictly follow this format:

[Reasoning regarding the five few-shot segments]

<Explain how the ground truth ratings in the prompt can be derived from the provided few-shot audio segments.>

[Summary Explaining the Ratings of the final test example]

<Detailed evidence log of the final test audio segment (the sixth segment), including detailed analysis of each sentence.>

<Reasoning for your rating of each evaluation metric>

<The rating of overall_score should take all factors into account.> \\

[Final Ratings]

expressive_intensity: <your rating>

expressive_correctness: <your rating>

intonation: <your rating>

nsvs_and_fillers: <your rating>

mispronunciation: <your rating>

pacing: <your rating>

overall_score: <your rating> \\

\# Important

- Please only judge the final conversation segment (the sixth segment), not the few-shot segments. The few-shot segments are only used as reference to help you understand the ground truth ratings. \\

\# Few-Shot Segment Examples

\#\# Conversation-1

\#\# Ratings:

\{Ground-truth Human Ratings\}

\#\# Conversation-2

\#\# Ratings:

\{Ground-truth Human Ratings\}

\#\# Conversation-3

\#\# Ratings:

\{Ground-truth Human Ratings\}

\#\# Conversation-4

\#\# Ratings:

\{Ground-truth Human Ratings\}

\#\# Conversation-5

\#\# Ratings:

\{Ground-truth Human Ratings\} \\

\# Now evaluate the final conversation segment in audio clip:
\end{tcolorbox}

\subsubsection{Few-shot Acoustic Feature Prompting.}
Instead of providing raw audio as context, we prompt a text-only LLM using explicitly extracted acoustic features. These features are obtained using the same vowel-level acoustic feature extraction pipeline described in Section~\ref{sec:gsrm} and illustrated in Figure~\ref{fig:teaser}. For each few-shot example, we concatenate the extracted acoustic features with the corresponding ground-truth human ratings, followed by the acoustic features of the test sample. The text LLM is then prompted to predict the ratings for the test sample. The prompt template used for acoustic feature prompting is shown below.
\begin{tcolorbox}[gray_box, title = {{Prompt Template: Few-shot Acoustic Feature Prompting}}]\footnotesize
You are an expert audio rater specializing in evaluating the quality of audio recordings by leveraging acoustic features.

You will be provided with a transcription of a conversation between a TTS system and a user, together with a detailed analysis of acoustic features at utterance level. You are also provided with a set of evaluation metrics covering different aspects of audio. 

Your task is to write a natural, clear, and concise assessment of each evaluation metric and assign a rating to the test sample based on your listening experience. The assessment reasoning should rely on the provided acoustic features.

Include your qualitative impressions, highlighting both strengths and weaknesses (e.g., drifts in pitch style across turns, or flatness) of the audio. Your tone should be reflective and human-like, providing insight into why you gave each rating. Each claim must be supported with evidence. \\

\# Evaluation Metrics

Each evaluation metric is described below:

- expressive_intensity: The intensity of expressiveness in the audio.

- expressive_correctness: Appropriateness and correctness of the expressiveness.

- intonation: Quality of the intonation in the speech.

- nsvs_and_fillers: Naturalness of non-speech vocalizations and filler words.

- mispronunciation: whether mispronunciation is present.

- pacing: Naturalness of the pacing or speed of the speech.

- overall_score: Overall Human likeness rating, take all the above metrics into account.

Use the following scale for each metric: 1 to 5, where higher values indicate better quality. Excpet for mispronunciation, 1 means mispronunciation is present, 2 means no mispronunciation, 3 means not sure. \\

\# Acoustic Features

<Detailed evidence log of the audio recording, including vowel-level analysis of each sentence.> \\

\# Instructions

When evaluating the audio, you must rely on the acoustic features—Pitch, Pitch Variation, Pitch Slope, Intensity, and Intensity Variation—as your primary evidence. Use them in the following expert-informed ways:

- Expressive Intensity: Identify the strength of expressiveness by looking for moderate to high Pitch Variation, clear Pitch Slopes, and sufficient Intensity. Very low variation in pitch or intensity indicates a flat, emotionless delivery. Extremely high variation may indicate unstable, exaggerated, or inconsistent expressiveness.

- Expressive Correctness: Evaluate whether the expressiveness is appropriate. Check if Pitch Variation and Intensity Variation are coherent and context-appropriate, not abrupt or mismatched. Excessively sharp slopes or sudden changes in loudness may suggest incorrect or unnatural expressiveness.

- Intonation Quality: Examine Pitch Slope to understand rising/falling contours, and Pitch Variation to assess natural melodic movement. Smooth slopes and moderate variation indicate good intonation. Flat pitch indicates monotone delivery; abrupt or irregular slopes indicate unstable intonation.

- NSVs and Fillers (Non-Speech Vocalizations): Look for sudden spikes in Intensity Variation or Pitch Variation that do not align with vowel structure. Irregular or abrupt changes may reflect throat noises, glottal catches, or filler-like disfluencies.

- Mispronunciation: Evaluate vowels for abnormally low intensity, very short or weak articulation (if duration provided), or unstable pitch contours. A vowel whose acoustic pattern deviates sharply from surrounding vowels may indicate mispronunciation.

- Pacing: Assess pacing consistency by checking whether the prosodic behavior across vowels is smooth and regular. Extreme or uneven Intensity Variation or Pitch. \\

\# Response Format

[Summary Explaining the Ratings]

<First check the above analysis for completeness and consistency; correct any errors; and add any missing issues or positives.>

<Reasoning for your rating of each evaluation metric. Refer to the details in the analysis.>

<The rating of overall_score should take all factors into account.> \\

[Final Ratings]

expressive_intensity: <your rating>

expressive_correctness: <your rating>

intonation: <your rating>

nsvs_and_fillers: <your rating>

mispronunciation: <your rating>

pacing: <your rating>

overall_score: <your rating> \\

\# Few-Shot Examples

\#\# Conversation-1

\#\# Acoustic Feature:

\{acoustic feature of sample-1\}

\#\# Ratings:

\{Ground-truth Human Ratings\}

\#\# Conversation-2

\#\# Acoustic Feature:

\{acoustic feature of sample-2\}

\#\# Ratings:

\{Ground-truth Human Ratings\}

\#\# Conversation-3

\#\# Acoustic Feature:

\{acoustic feature of sample-3\}

\#\# Ratings:

\{Ground-truth Human Ratings\}

\#\# Conversation-4

\#\# Acoustic Feature:

\{acoustic feature of sample-4\}

\#\# Ratings:

\{Ground-truth Human Ratings\}

\#\# Conversation-5

\#\# Acoustic Feature:

\{acoustic feature of sample-5\}

\#\# Ratings:

\{Ground-truth Human Ratings\} \\

\# Now evaluate the final conversation:

\#\# Acoustic Feature:

\{acoustic feature of test sample\}
\end{tcolorbox}

\subsubsection{MOS Predictors.}
We include existing utterance-level MOS predictors as baselines, specifically NISQA\footnote{The NISQA model code used in this research is licensed under the MIT License. The model weights are licensed under the Creative Commons Attribution-NonCommercial-ShareAlike 4.0 International (CC BY-NC-SA 4.0) License. The weights were used solely for non-commercial research purposes in accordance with the terms of this license. No distribution of the model weights or code containing the weights is included in this publication.}~\cite{mittag2021nisqa} and UTMOSv2~\cite{baba2024utmosv2}. Both models are originally designed for single-speaker, single-utterance speech naturalness assessment and output a scalar MOS score. Since these predictors do not support multi-turn conversational inputs, we apply them independently to each utterance within a conversation. For example, if a conversation consists of five utterances, the predictor produces five corresponding MOS scores. Conversation-level predictions are then obtained by aggregating the utterance-level scores using simple statistics, such as the mean or minimum.

\subsubsection{Regression-based Predictors.}
In contrast to MOS predictors, regression-based predictors are explicitly trained to map conversational speech representations to human-annotated MOS scores. These models follow a unified regression framework built on top of large pre-trained speech encoders, including AES- and WavLM-based backbones. Given an input conversation, the audio is segmented into fixed-length windows, and frame-level representations are extracted using a frozen transformer encoder. A weighted sum of hidden states across all encoder layers is computed, followed by temporal mean pooling to obtain a segment-level embedding. The resulting embeddings are then passed through a multi-layer perceptron with GeLU activations and dropout, trained with mean squared error loss to predict MOS scores. Final conversation-level predictions are obtained by averaging the regression outputs across all segments.

\subsection{Implementation Details of GSRM}

\subsubsection{Acoustic Feature Extraction.}
\label{app:acoustic_features}

Given a speech recording and its transcript, all audio is first resampled to 24 kHz. For multi-channel recordings, only the first channel is retained (system response). Extended silences are removed using a Silero-based voice activity detector with a merge threshold of 1000 ms to preserve natural intra-utterance timing.

Phoneme-level forced alignment is then performed between the transcript and waveform using a neural alignment model. From the resulting alignment, we retain only vowel segments. The vowel inventory is defined using an IPA-based set covering both monophthongs and diphthongs. For each aligned vowel, two temporal spans are defined. The \emph{full span}, computed from the earliest onset to the latest offset across all alignment states, is used to measure vowel duration. The \emph{core span} is a stable sub-region used for pitch and intensity estimation to avoid boundary transitions. When five or more alignment states are available, the middle two states are selected; when three to four states are present, the central state is used; otherwise, 20\% of the full span is trimmed from each edge.

From each vowel segment, we extract a compact set of low-level prosodic features: (i) pitch level, defined as the mean fundamental frequency (F0) over the core span; (ii) pitch variation, computed as the standard deviation of F0; (iii) pitch slope, obtained via a linear fit to F0 over time; (iv) intensity level, measured as mean RMS energy; (v) intensity variation, defined as the standard deviation of RMS energy; and (vi) duration, defined as the length of the full span. Pitch extraction follows a Praat-style autocorrelation method with speaker-adaptive floor and ceiling parameters, and all pitch and intensity statistics are computed only over voiced frames.

Continuous feature values are normalized using a two-stage procedure. First, features are z-normalized at the speaker level to control for individual vocal characteristics. Second, vowel-type normalization is applied to mitigate systematic differences across vowel classes. The normalized values are then discretized into ordinal categories. Specifically, pitch level is discretized using fixed thresholds at 85, 110, 160, 220, 280, and 390~Hz, corresponding to \emph{extremely low}, \emph{very low}, \emph{low}, \emph{mid}, \emph{high}, \emph{very high}, and \emph{extremely high}. Pitch slope is categorized separately for rising and falling contours using the 25th and 75th percentiles of positive and negative slopes, yielding \emph{low}, \emph{mid}, and \emph{high} slope categories. Pitch variation, intensity level, intensity variation, and duration are discretized via quantile-based binning at the 10th, 25th, 75th, and 90th percentiles, producing five ordinal categories: \emph{very low}, \emph{low}, \emph{mid}, \emph{high}, and \emph{very high}. Duration quantiles are computed only from vowels with fully voiced cores.

The final output is an utterance-level acoustic feature log that records vowel sequences and their discretized prosodic attributes in temporal order. We provide an example of the extracted vowel-level acoustic feature log below.

\begin{tcolorbox}[blue_box, title = {{Example: Vowel-level Acoustic Feature}}]\footnotesize
\#\# Transcript: ``so there could be a few reasons why he's doing this."

\#\#\# The vowel /\IPA{o}/ in ``so": [Pitch] mid, [Pitch Variation] low, [Pitch Slope] high falling, [Intensity] high, [Intensity variation] mid, [Duration] very high.

\#\#\# The vowel /\IPA{ɛ}/ in ``there": [Pitch] mid, [Pitch Variation] mid, [Pitch Slope] high falling, [Intensity] high, [Intensity variation] very low, [Duration] low.

\#\#\# The vowel /\IPA{ʊ}/ in ``could": [Pitch] high, [Pitch Variation] very high, [Pitch Slope] high falling, [Intensity] mid, [Intensity variation] mid, [Duration] low.

\#\#\# The vowel /\IPA{iː}/ in ``be": [Pitch] high, [Pitch Variation] mid, [Pitch Slope] high falling, [Intensity] mid, [Intensity variation] low, [Duration] mid.

\#\#\# The vowel /\IPA{ə}/ in ``a": [Pitch] high, [Pitch Variation] high, [Pitch Slope] high falling, [Intensity] high, [Intensity variation] mid, [Duration] low.

\#\#\# The vowel /\IPA{uː}/ in ``few": [Pitch] mid, [Pitch Variation] mid, [Pitch Slope] high falling, [Intensity] mid, [Intensity variation] mid, [Duration] mid.

\#\#\# The vowel /\IPA{iː}/ in ``reasons": [Pitch] mid, [Pitch Variation] mid, [Pitch Slope] high falling, [Intensity] mid, [Intensity variation] low, [Duration] mid.

\#\#\# The vowel /\IPA{a}/ in ``why": [Pitch] low, [Pitch Variation] mid, [Pitch Slope] high falling, [Intensity] mid, [Intensity variation] very low, [Duration] low.

\#\#\# The vowel /\IPA{iː}/ in ``he's": [Pitch] low, [Pitch Variation] mid, [Pitch Slope] high falling, [Intensity] low, [Intensity variation] mid, [Duration] mid.

\#\#\# The vowel /\IPA{uː}/ in ``doing": [Pitch] low, [Pitch Variation] mid, [Pitch Slope] high falling, [Intensity] mid, [Intensity variation] low, [Duration] mid.

\#\#\# The vowel /\IPA{ɪ}/ in ``this": [Pitch] low, [Pitch Variation] mid, [Pitch Slope] high falling, [Intensity] low, [Intensity variation] mid, [Duration] high.

\#\# Transcript: ``he might be seeking validation from his friends, or maybe he's not aware of how his behavior affects you."

\#\#\# The vowel /\IPA{iː}/ in ``he": [Pitch] high, [Pitch Variation] low, [Pitch Slope] low rising, [Intensity] mid, [Intensity variation] mid, [Duration] low.

\#\#\# The vowel /\IPA{a}/ in ``might": [Pitch] high, [Pitch Variation] mid, [Pitch Slope] mid rising, [Intensity] mid, [Intensity variation] very low.

\#\#\# The vowel /\IPA{iː}/ in ``be": [Pitch] high, [Pitch Variation] high, [Pitch Slope] high falling, [Intensity] mid, [Intensity variation] low, [Duration] low.

\#\#\# The vowel /\IPA{iː}/ in ``seeking": [Pitch] mid, [Pitch Variation] high, [Pitch Slope] high falling, [Intensity] mid, [Intensity variation] high, [Duration] mid.

\#\#\# The vowel /\IPA{æ}/ in ``validation": [Pitch] mid, [Pitch Variation] mid, [Pitch Slope] high falling, [Intensity] mid, [Intensity variation] low, [Duration] mid.

\#\#\# The vowel /\IPA{ɪ}/ in ``his": [Pitch] low, [Pitch Variation] mid, [Pitch Slope] high falling, [Intensity] low, [Intensity variation] mid, [Duration] low.

\#\#\# The vowel /\IPA{ɛ}/ in ``friends": [Pitch] low, [Pitch Variation] mid, [Pitch Slope] high falling, [Intensity] low, [Intensity variation] mid, [Duration] high.

\#\#\# The vowel /\IPA{ɔː}/ in ``or": [Pitch] low, [Pitch Variation] mid, [Pitch Slope] high falling, [Intensity] low, [Intensity variation] very high.

\#\#\# The vowel /\IPA{e}/ in ``maybe": [Pitch] low, [Pitch Variation] low, [Pitch Slope] high falling, [Intensity] low, [Intensity variation] very low, [Duration] mid.

\#\#\# The vowel /\IPA{iː}/ in ``maybe": [Pitch] low, [Pitch Variation] mid, [Pitch Slope] high falling, [Intensity] low, [Intensity variation] mid, [Duration] mid.

\#\#\# The vowel /\IPA{iː}/ in ``he's": [Pitch] low, [Pitch Variation] mid, [Pitch Slope] high falling, [Intensity] low, [Intensity variation] mid, [Duration] mid.

\#\#\# The vowel /\IPA{ɒ}/ in ``not": [Pitch] low, [Pitch Variation] mid, [Pitch Slope] high falling, [Intensity] low, [Intensity variation] very low, [Duration] mid.

\#\#\# The vowel /\IPA{ə}/ in ``aware": [Pitch] low, [Pitch Variation] low, [Pitch Slope] high falling, [Intensity] low, [Intensity variation] mid, [Duration] mid.

\#\#\# The vowel /\IPA{a}/ in ``how": [Pitch] low, [Pitch Variation] mid, [Pitch Slope] high falling, [Intensity] low, [Intensity variation] high, [Duration] mid.

\#\#\# The vowel /\IPA{ɪ}/ in ``his": [Pitch] low, [Pitch Variation] low, [Pitch Slope] mid rising, [Intensity] low, [Intensity variation] low, [Duration] mid.

\#\#\# The vowel /\IPA{ɪ}/ in ``behavior": [Pitch] low, [Pitch Variation] mid, [Pitch Slope] high falling, [Intensity] low, [Intensity variation] mid, [Duration] mid.

\#\#\# The vowel /\IPA{uː}/ in ``you": [Pitch] low, [Pitch Variation] mid, [Pitch Slope] high falling, [Intensity] very low, [Intensity variation] very high.

\#\# Transcript: ``he could be trying to assert dominance or control, or maybe he's got some insecurity issues."

\#\#\# The vowel /\IPA{iː}/ in ``he": [Pitch] high, [Pitch Variation] mid, [Pitch Slope] high falling, [Intensity] mid, [Intensity variation] high, [Duration] mid.

\#\#\# The vowel /\IPA{ʊ}/ in ``could": [Pitch] high, [Pitch Variation] mid, [Pitch Slope] high falling, [Intensity] mid, [Intensity variation] mid, [Duration] low.

\#\#\# The vowel /\IPA{iː}/ in ``be": [Pitch] high, [Pitch Variation] mid, [Pitch Slope] high falling, [Intensity] mid, [Intensity variation] mid, [Duration] mid.

\#\#\# The vowel /\IPA{a}/ in ``trying": [Pitch] mid, [Pitch Variation] mid, [Pitch Slope] high falling, [Intensity] mid, [Intensity variation] low, [Duration] low.

\#\#\# The vowel /\IPA{ə}/ in ``to": [Pitch] mid, [Pitch Variation] mid, [Pitch Slope] high falling, [Intensity] mid, [Intensity variation] mid, [Duration] low.

\#\#\# The vowel /\IPA{ə}/ in ``assert": [Pitch] mid, [Pitch Variation] mid, [Pitch Slope] high falling, [Intensity] mid, [Intensity variation] mid, [Duration] mid.

\#\#\# The vowel /\IPA{ɒ}/ in ``dominance": [Pitch] mid, [Pitch Variation] mid, [Pitch Slope] high falling, [Intensity] mid, [Intensity variation] low, [Duration] high.

\#\#\# The vowel /\IPA{ə}/ in ``control": [Pitch] low, [Pitch Variation] mid, [Pitch Slope] high falling, [Intensity] mid, [Intensity variation] low, [Duration] low.

\#\#\# The vowel /\IPA{ɔː}/ in ``or": [Pitch] mid, [Pitch Variation] mid, [Pitch Slope] low rising, [Intensity] very low, [Intensity variation] very high.

\#\#\# The vowel /\IPA{ɒ}/ in ``got": [Pitch] mid, [Pitch Variation] high, [Pitch Slope] high falling, [Intensity] mid, [Intensity variation] mid, [Duration] mid.

\#\#\# The vowel /\IPA{ʌ}/ in ``some": [Pitch] mid, [Pitch Variation] mid, [Pitch Slope] high falling, [Intensity] mid, [Intensity variation] low, [Duration] low.

\#\#\# The vowel /\IPA{ɪ}/ in ``insecurity": [Pitch] mid, [Pitch Variation] very low, [Pitch Slope] high falling, [Intensity] mid, [Intensity variation] very low, [Duration] low.

\#\#\# The vowel /\IPA{ɪ}/ in ``issues": [Pitch] low, [Pitch Variation] mid, [Pitch Slope] high falling, [Intensity] mid, [Intensity variation] low, [Duration] low.

\#\# Transcript: ``it's also possible he's not considering your feelings."

\#\#\# The vowel /\IPA{iː}/ in ``he's": [Pitch] high, [Pitch Variation] mid, [Pitch Slope] high falling, [Intensity] mid, [Intensity variation] low, [Duration] mid.

\#\#\# The vowel /\IPA{ɪ}/ in ``it's": [Pitch] mid, [Pitch Variation] very low, [Pitch Slope] low rising, [Intensity] very low, [Intensity variation] very high.

\#\#\# The vowel /\IPA{ɒ}/ in ``possible": [Pitch] mid, [Pitch Variation] high, [Pitch Slope] high falling, [Intensity] mid, [Intensity variation] high, [Duration] high.

\#\#\# The vowel /\IPA{iː}/ in ``he's": [Pitch] mid, [Pitch Variation] mid, [Pitch Slope] high falling, [Intensity] mid, [Intensity variation] mid, [Duration] high.

\#\#\# The vowel /\IPA{ɒ}/ in ``not": [Pitch] mid, [Pitch Variation] low, [Pitch Slope] high falling, [Intensity] mid, [Intensity variation] high, [Duration] high.

\#\#\# The vowel /\IPA{ə}/ in ``considering": [Pitch] mid, [Pitch Variation] mid, [Pitch Slope] high falling, [Intensity] mid, [Intensity variation] mid.

\#\#\# The vowel /\IPA{iː}/ in ``feelings": [Pitch] mid, [Pitch Variation] low, [Pitch Slope] high falling, [Intensity] mid, [Intensity variation] low, [Duration] high.

\#\# Transcript: ``have you talked to him about how you feel when this happens?"

\#\#\# The vowel /\IPA{æ}/ in ``have": [Pitch] mid, [Pitch Variation] low, [Pitch Slope] high falling, [Intensity] mid, [Intensity variation] mid, [Duration] mid.

\#\#\# The vowel /\IPA{ɔː}/ in ``talked": [Pitch] high, [Pitch Variation] mid, [Pitch Slope] high falling, [Intensity] mid, [Intensity variation] mid, [Duration] mid.

\#\#\# The vowel /\IPA{ɪ}/ in ``him": [Pitch] high, [Pitch Variation] mid, [Pitch Slope] high falling, [Intensity] mid, [Intensity variation] very low.

\#\#\# The vowel /\IPA{ə}/ in ``about": [Pitch] mid, [Pitch Variation] high, [Pitch Slope] high falling, [Intensity] mid, [Intensity variation] mid, [Duration] low.

\#\#\# The vowel /\IPA{a}/ in ``how": [Pitch] mid, [Pitch Variation] mid, [Pitch Slope] high falling, [Intensity] mid, [Intensity variation] mid, [Duration] mid.

\#\#\# The vowel /\IPA{uː}/ in ``you": [Pitch] mid, [Pitch Variation] mid, [Pitch Slope] high falling, [Intensity] low, [Intensity variation] mid, [Duration] mid.

\#\#\# The vowel /\IPA{iː}/ in ``feel": [Pitch] mid, [Pitch Variation] mid, [Pitch Slope] high falling, [Intensity] mid, [Intensity variation] low, [Duration] mid.

\#\#\# The vowel /\IPA{ɛ}/ in ``when": [Pitch] mid, [Pitch Variation] mid, [Pitch Slope] high falling, [Intensity] mid, [Intensity variation] very low, [Duration] low.

\#\#\# The vowel /\IPA{ɪ}/ in ``this": [Pitch] low, [Pitch Variation] mid, [Pitch Slope] high falling, [Intensity] low, [Intensity variation] mid, [Duration] low.

\#\#\# The vowel /\IPA{æ}/ in ``happens": [Pitch] mid, [Pitch Variation] mid, [Pitch Slope] high falling, [Intensity] low, [Intensity variation] mid, [Duration] mid.

\end{tcolorbox}

\subsubsection{Reasoning Synthesis.}
After extracting the acoustic feature log for an audio sample, we leverage a teacher model, GPT-4o, to synthesize the utterance-level evidence log. Specifically, for each utterance (sentence) and its associated vowel-level acoustic features, we prompt GPT-4o to generate a structured analysis capturing the inferred context, perceptual strengths, and potential issues of the utterance. We then concatenate the analyses of all utterances to form the complete evidence log for the audio sample. The prompt template used for evidence log synthesis is provided below.
\begin{tcolorbox}[gray_box, title = {{Prompt Template: Synthesizing Utterence-level Evidence Log}}]\footnotesize
You are an expert audio rater specializing in evaluating the quality of audio recordings. Your are given,

- A transcript of a spoken sentence.

- Acoustic features describing each vowel in the sentence.

- Evaluation metrics you must judge.

Your goal is to produce an analysis:

[Inferred context]: 

question/empathetic/apology/celebration/instruction/neutral statement. If no clear context exists, say: “No clear context inferred.”

[NSVs/fillers]: 

List any non-speech vocalizations (NSVs) or filler words (laugh, yeah, Aww, uh, hmm, etc.) that are present in the audio. If none exist, say: “No NSVs/fillers detected.”

[Positive]: 

Highlight the most positive aspects of the audio: expressive intensity, good emphasis, varied intonation, natural fillers/laughs, pleasant timbre. If no clear positive aspects exist, say: “No positive aspects detected.”

[Potential Issue]: 

Point out any potential issues: sharp pitch changes, intonation problems, pacing issues, speaker style changes, expressive incorrectness, mispronunciation, flat intonation, unnatural pauses, or any other issues. If no clear issue exists, say: “No notable issues detected.” \\

\# Evaluation Metrics

Each evaluation metric is described below:

- expressive_intensity: The intensity of expressiveness in the audio.

- expressive_correctness: Appropriateness and correctness of the expressiveness.

- intonation: Quality of the intonation in the speech.

- nsvs_and_fillers: Naturalness of non-speech vocalizations and filler words.

- mispronunciation: whether mispronunciation is present.

- pacing: Naturalness of the pacing or speed of the speech. \\

\# The system responses and acoustic features:

\{Acoustic features\} \\

\# Instructions

When evaluating the audio, you must rely on the provided acoustic features—Pitch, Pitch Variation, Pitch Slope, Intensity, and Intensity Variation—as your primary evidence. 

Use them in the following expert-informed ways:

- Expressive Intensity: Identify the strength of expressiveness by looking for moderate to high Pitch Variation, clear Pitch Slopes, and sufficient Intensity. Very low variation in pitch or intensity indicates a flat, emotionless delivery. Extremely high variation may indicate unstable, exaggerated, or inconsistent expressiveness.

- Expressive Correctness: Evaluate whether the expressiveness is appropriate. Check if Pitch Variation and Intensity Variation are coherent and context-appropriate, not abrupt or mismatched. Excessively sharp slopes or sudden changes in loudness may suggest incorrect or unnatural expressiveness.

- Intonation Quality: Examine Pitch Slope to understand rising/falling contours, and Pitch Variation to assess natural melodic movement. Smooth slopes and moderate variation indicate good intonation. Flat pitch indicates monotone delivery; abrupt or irregular slopes indicate unstable intonation.

- NSVs and Fillers (Non-Speech Vocalizations): Look for sudden spikes in Intensity Variation or Pitch Variation that do not align with vowel structure. Irregular or abrupt changes may reflect throat noises, glottal catches, or filler-like disfluencies.

- Mispronunciation: Evaluate vowels for abnormally low intensity, very short or weak articulation (if duration provided), or unstable pitch contours. A vowel whose acoustic pattern deviates sharply from surrounding vowels may indicate mispronunciation.

- Pacing: Assess pacing consistency by checking whether the prosodic behavior across vowels is smooth and regular. Extreme or uneven Intensity Variation or Pitch Variation may reflect rushed or unnatural pacing. If duration is given, long durations imply slow pacing; short durations imply fast pacing. \\

\# Response Format

Your response must strictly follow this format:

[Inferred context]

<one concise sentence describing the inferred context, referencing the transcripts as evidence.> OR ``No clear context inferred."

[NSVs/fillers]

<one concise sentence listing any NSVs/fillers, referencing the transcripts and features as evidence.> OR ``No NSVs/fillers detected."

[Positive]

<one concise sentence highlighting positive aspects of the evaluation metrics, referencing the features as evidence.> OR ``No positive aspects detected."

[Potential Issue]

<one concise sentence describing weaknesses, referencing the features as evidence.> OR ``No notable issues detected." \\

\# Important

- Always refer specifically to the words and vowels in the transcript as evidence. Do Not say something too generic without referencing the features.

- Although always use features and transcripts as evidence, Do Not explicitly mention something like ``Based on the features and transcripts, I judge the audio as follows:" in your response.
\end{tcolorbox}

Once the evidence log is constructed, we prompt GPT-4o again to synthesize a global judgment chain-of-thought (CoT). We adopt a \emph{role-playing} prompt that grants the model access to both the evidence log and the ground-truth human ratings, and instructs it to explain the reasoning behind the assigned scores. This design ensures that the synthesized reasoning is faithful to the oracle ratings while remaining coherent. The prompt template for global judgment synthesis is shown below.
\begin{tcolorbox}[gray_box, title = {{Prompt Template: Synthesizing Global Judgment CoT}}]\footnotesize
You are an audio assessment assistant striving to become an expert audio rater specializing in evaluating the quality of audio recordings. You are now taking an exam to evaluate your capabilities. 

You will be provided with a transcription of a conversation between a TTS system and a user, together with a detailed analysis of acoustic features at sentence level (user turns are ommitted). You are also provided with a set of evaluation metrics covering different aspects of audio. 

Your task is to write a natural, clear, and concise assessment of each evaluation metric and assign a rating based on your listening experience. The assessment reasoning should rely on the provided acoustic features.

Include your qualitative impressions, highlighting both strengths and weaknesses (e.g., drifts in pitch style across turns, or flatness) of the audio. Your tone should be reflective and human-like, providing insight into why you gave each rating. Each claim must be supported with evidence.

To evaluate your correctness, the oracle ratings will also be provided. You must ensure that your ratings MATCH the oracle ratings EXACTLY. 
However, your reasoning process must appear fully self-derived and must NOT reference, suggest awareness of, or appear to be influenced by the oracle ratings. \\

\# Evaluation Metrics

Each evaluation metric is described below:

- expressive_intensity: The intensity of expressiveness in the audio.

- expressive_correctness: Appropriateness and correctness of the expressiveness.

- intonation: Quality of the intonation in the speech.

- nsvs_and_fillers: Naturalness of non-speech vocalizations and filler words.

- mispronunciation: whether mispronunciation is present.

- pacing: Naturalness of the pacing or speed of the speech.

- overall_score: Overall Human likeness rating \\

\# Oracle Ratings (For Evaluation Only):

Use the following scale for each metric: 1-5. Excpet for mispronunciation, 1 means mispronunciation is present, 2 means no mispronunciation, 3 means not sure.

expressive_intensity: {expressive_intensity}

expressive_correctness: {expressive_correctness}

intonation: {intonation}

nsvs_and_fillers: {nsvs_and_fillers}

mispronunciation: {mispronunciation}

pacing: {pacing}

overall_score: {human_likeness_speaker_2} \\

\# The system responses and detailed analysis of acoustic features:

\{Evidence Log\} \\

\# Response Format 

Your response must strictly follow this format:

[Summary Explaining the Ratings]

<First check the above analysis for completeness and consistency; correct any errors; and add any missing issues or positives.>

<Reasoning for your rating of each evaluation metric. Refer to the details in the analysis.>

<The rating of overall_score should take all factors into account.> \\

[Final Ratings]

expressive_intensity: <your rating>

expressive_correctness: <your rating>

intonation: <your rating>

nsvs_and_fillers: <your rating>

mispronunciation: <your rating>

pacing: <your rating>

overall_score: <your rating>
\end{tcolorbox}

\subsubsection{Training and Inference.}
We fine-tune GSRM on the ConvTTS training set using Qwen2.5-Omni-7B as the backbone model. Given an input speech recording, GSRM is trained to first generate the utterance-level evidence log, followed by the global judgment CoT and the final rating scores. The supervised fine-tuning (SFT) prompt template used for training GSRM is provided below.
\begin{tcolorbox}[gray_box, title = {{Prompt Template: GSRM Training \& Inference }}]\footnotesize
You are an expert audio rater specializing in evaluating the quality of audio recordings.

You will be provided with an audio recording of a conversation between a TTS system and a user. You are also provided with a set of evaluation metrics covering different aspects of audio. 

Your task is to write a natural, clear, and concise assessment of each evaluation metric and assign a rating based on your listening experience.

Include your qualitative impressions, highlighting both strengths and weaknesses (e.g., drifts in pitch style across turns, or flatness) of the audio. 

Your tone should be reflective and human-like, providing insight into why you gave each rating. Each claim must be supported with evidence from the audio recording. \\

\# Evaluation Metrics

Each evaluation metric is described below:

- expressive_intensity: The intensity of expressiveness in the audio.

- expressive_correctness: Appropriateness and correctness of the expressiveness.

- intonation: Quality of the intonation in the speech.

- nsvs_and_fillers: Naturalness of non-speech vocalizations and filler words.

- mispronunciation: whether mispronunciation is present.

- pacing: Naturalness of the pacing or speed of the speech.

- overall_score: Overall Human likeness rating \\

\# Instruction and Response Format

Please follow these steps to complete the task. Your response must strictly follow this format: \\

[Evidence log]

<Detailed evidence log of the audio recording, including vowel-level analysis of each sentence.> \\

[Summary Explaining the Ratings]

<Reasoning for your rating of each evaluation metric>

<The rating of overall_score should take all factors into account.> \\

[Final Ratings]

expressive_intensity: <your rating>

expressive_correctness: <your rating>

intonation: <your rating>

nsvs_and_fillers: <your rating>

mispronunciation: <your rating>

pacing: <your rating>

overall_score: <your rating>
\end{tcolorbox}

We use a learning rate of $2\times10^{-5}$ with a cosine learning rate scheduler. The batch size is set to 32, the maximum sequence length is 4096, and the model is trained for up to 10 epochs. During inference, we select the model checkpoint with the lowest validation loss on a held-out validation set. For each audio sample, we perform up to 16 random inference passes with a sampling temperature of 1.0 and top-$p$ set to 0.6. The final rating is obtained by averaging the scores across these samples. All experiments are conducted using the SWIFT framework~\cite{zhao2024swiftascalablelightweightinfrastructure}.

\section{Details of Online RLHF for Speech LLM} \label{sec:rlhf_gtrm}

\subsection{GSRM for Speech Semantic Judge} \label{sec:semantic_judge}
\begin{table*}[h]
\centering
\footnotesize
\renewcommand{\arraystretch}{1}  
\setlength{\tabcolsep}{10pt}
\captionsetup{font=small}

\caption{\textbf{Annotation Rubric for Speech Semantic Quality.}
Each dialog transcript is evaluated along multiple sub-dimensions that capture different aspects of speech semantic quality.}
\vspace{-0.6em}

\begin{tabular}{p{4.3cm} p{11.4cm}}
\toprule
\textbf{Sub-metric} & \textbf{Description} \\
\midrule
\textbf{Language Complexity} & For language complexity, consider two main aspects: \textbf{1. Vocabulary \& Wording} -- Favor responses that use colloquial language, idiomatic expressions, contractions (e.g., “can’t”, “it’s”), deixis (e.g., “this”, “that”, “here”), and, where appropriate, mild slang. Penalize responses that use overly formal, esoteric, or academic vocabulary, as well as robotic or non-idiomatic phrasing. \textbf{2. Syntax} -- Favor simple, straightforward sentence structures. Look for short sentences, frequent pauses, and incremental phrasing (e.g., “Yeah, and then…”), which contribute to a natural rhythm and pacing. Penalize sentences that are too long without breaks, or that use unnecessarily complex syntax. Consider the following questions --- Does the assistant’s response sound like something a real person would say in a casual spoken conversation? Is the language accessible and easy to follow in verbal conversation? Are there any words, phrases, or sentence structures that feel unnatural, overly formal, or robotic for spoken dialog? \\
\textbf{Contextual Awareness} & Consider three main aspects: \textbf{1. Contents} -- Favor responses that are straightforward and direct, frontloading key information. Penalize responses that are off-topic, irrelevant, circuitous, indirect, or simply repeat the user’s query. Adjust for context: In emergencies or urgent situations, concise and directive speech is preferred. In teaching scenarios or when the user requests details, more elaborate explanations are appropriate. \textbf{2. Length} -- Favor brief responses that are clear and complete, while avoiding unnecessary terseness. Penalize responses that are long-winded or contain unnecessary elaboration. \textbf{3. Tone} -- Favor a friendly, approachable, and engaging tone by default. Adjust tone based on user cues: If the user is upset, the assistant should soften its tone and convey understanding or empathy. If the user is excited, the assistant should respond with enthusiasm. Always match the tone to the context and user sentiment \\
\textbf{Spontaneity} & A response is considered \textbf{spontaneous} if it includes one or more of the following features, which are typical of authentic, on-the-fly human speech (as long as they do not significantly harm clarity or delivery, e.g., by making the response confusing or incoherent): \textbf{Frequent checks:} Phrases like “you know?”, “right?”, or similar, used to engage or check understanding. \textbf{Self-repairs:} Corrections or adjustments to previous statements, e.g., “I mean...”. \textbf{Clarifications:} Paraphrasing, moderate repetition, or rephrasing to ensure the message is easy to understand. \textbf{Disfluency:} Natural filler words (“like”, “well”, “hmm”, "so"), restarts, or false starts. A response is considered \textbf{non-spontaneous} if it lacks these features and instead sounds polished, rehearsed, or as if being read from a script.\\
\bottomrule
\end{tabular}
\label{table:semantic_rubric}
\end{table*}

We extend GSRM to evaluate semantic aspects of speech naturalness across dimensions such as language complexity, contextual awareness, and spontaneity, as detailed in Table \ref{table:semantic_rubric}. A model’s response may sound aesthetically natural—with near-perfect pacing and intonation—yet still fail semantically, for example by ignoring conversational context, responding with inappropriate affect (e.g., being overly cheerful or rude to a sad query), or drifting completely off-topic. Similarly, responses that are excessively verbose or overly formal can undermine both spontaneity and the perceived naturalness of language complexity. Importantly, these failure modes can be identified purely from the text transcript.

\textbf{Label and CoT Generation.} In contrast to Section~\ref{sec:human_data} where GSRM for audio naturalness required human annotations for dialogue-level dimension scores, we use advanced text-only reasoning models that are carefully prompted to produce binary scores for each dimension of semantic naturalness. We initially experimented with several frontier models, including Gemini 2.5 Pro, GPT-4o, and Llama 3.1, and ensembled their predictions to derive a final dimension-level score for each dialogue. However, we ultimately converged on GPT-OSS-120B \citep{agarwal2025gpt}, which produced labels that best matched our internally curated golden labels across dimensions.

To reduce variance in the judge’s labels and approximate multi-rater human evaluation, we sample 10 independent judgments per dialogue per dimension and then take the majority vote as the final dimension score. Rather than simply summing these dimension scores into a single naturalness metric, we further prompt GPT-OSS-120B to consider the oracle per-dimension scores and the dialogue context to produce a final semantic naturalness score between 0 and 3 along with a CoT reasoning as shown in the prompt template below.  

\begin{tcolorbox}[gray_box, title = {{Prompt Template: GSRM for Speech Semantic Quality Training \& Inference }}]\footnotesize
You are a semantic quality evaluation expert tasked with assessing the assistant's responses in a user-assistant conversation. Evaluate the assistant based on 3 criteria: Spontaneity, Language Simplicity, and Context Awareness.\\

[Instructions]

1. Review the conversation between the user and the assistant.
2. For each criterion, provide a score (0 or 1) and a concise reasoning (<=50 words) referencing specific assistant responses.\\

\# Evaluation Metrics

1. Spontaneity: A response is considered spontaneous if it includes features typical of authentic human speech, such as:

- Frequent checks (e.g., "you know?", "right?"),

- Self-repairs (e.g., "I mean..."),        

- Clarifications (e.g., paraphrasing, moderate repetition),

- Disfluency (e.g., natural filler words, restarts, or false starts).

A response is non-spontaneous if it lacks these features and sounds polished, rehearsed, or scripted.\\

2. Language Complexity: Evaluate the language complexity based on two aspects:

- Vocabulary \& Wording: Favor colloquial language, idiomatic expressions, contractions, and mild slang. Penalize overly formal, esoteric, or academic vocabulary, and robotic or non-idiomatic phrasing.

- Syntax: Favor simple, straightforward sentence structures, short sentences, and incremental phrasing. Penalize long, complex sentences without breaks. Consider if the response sounds like natural spoken conversation and is easy to follow.\\

3. Context Awareness: Assess if the assistant demonstrates understanding of the user's intent and situation, and adapts its response accordingly in:

- Contents: Favor direct, relevant, and concise responses, adjusting for context (e.g., emergencies, teaching scenarios).

- Length: Favor brief, clear, and complete responses, avoiding unnecessary elaboration. 

- Tone: Favor a friendly, approachable tone, adjusting based on user cues (e.g., upset, excited).\\

[Output Format]

\{\{

"spontaneity_score": <0 or 1>,

"spontaneity_reasoning": "< <=50 words>",

"language_complexity_score": <0 or 1>,

"language_complexity_reasoning": "< <=50 words>",

"context_awareness_score": <0 or 1>,

"context_awareness_reasoning": "< <=50 words>",

"overall_score": <0-3>,

"analysis": "< <=50 words summary referencing assistant responses and criteria>"

\}\}\\

\# Response Format

Assign an overall score (0-3) based on the sub-dimension scores and other aspects of the assistant's response. \textit{Consider the sub-dimension scores as a guide, but also look for other cues in the assistant's response that may not be captured by these criteria. A score of 3 indicates a perfect human-like conversation transcript, while a score of 0 indicates a response that is clearly robotic, inappropriate, or scripted.} Use your judgment to weigh the importance of each sub-dimension score and other factors in determining the overall score.\\

\# Oracle Ratings

You will be provided with a conversation between a user and an assistant, along with oracle ratings for the evaluation criteria. Ensure that your ratings match the oracle ratings \textit{exactly} for the sub-dimension scores, without referencing or alluding to them in your reasoning.

\{\{

"oracle_spontaneity_score": \{oracle_spontaneity_score\},  

"oracle_language_complexity_score": \{oracle_language_complexity_score\},   

"oracle_contextual_awareness_score": \{oracle_contextual_awareness_score\}

\}\}\\

Remember: Your response should be a JSON object; do not say anything else. Begin your response with a '\{\{' and end with a '\}\}'.

---

Now evaluate the following dialog: \{dialog\}"
\end{tcolorbox}

\textbf{Training and Evaluation.}
\label{sec: gsrm training} We fine-tuned a Qwen-2.5-Omni-7B model to predict sub-dimension scores, the overall naturalness score, and chain-of-thought (CoT) reasoning traces from user-assistant dialogue transcriptions. As an ablation to understand the impact of including CoT reasoning, we also trained a model to directly predict the overall naturalness score without generating CoT traces, similar to direct score predictor in Section~\ref{sec:pilot_studies}.

Labels and CoT reasoning traces were generated in-house on samples produced by an ensemble of different audio foundation models, using audio prompts designed to elicit emotional responses. Our training dataset consisted of 11,000 such samples. In-domain (on-policy) evaluation was performed on 100 samples from the same distribution as the training set, while for out-of-domain (OOD) evaluation, we collected labels on 130 samples from the FDX-Conv dataset described in Section~\ref{sec:human_data}.

All models were trained for 250 steps with a batch size of 512. The model trained with chain-of-thought (CoT) reasoning converged rapidly, reaching convergence within 95 steps. In contrast, the direct score predictor baseline required approximately 200 steps to converge. As will be discussed in Section \ref{sec: gtrm results}, incorporating CoT reasoning leads to faster convergence, substantially improved generalization, and stronger correlation with ground-truth semantic labels. These results highlight the effectiveness of framing semantic reward modeling as a generative task.

\subsection{Applying GSRM to Online RLHF} \label{sec:rlhf_details}
\textbf{Base Model and RLHF Data.} The base model is an in-house full-duplex speech LLM capable of listening and speaking simultaneously. While the user is speaking, the model can continuously generate spoken responses while understanding the user’s input. The model is trained on a large-scale speech corpus through a pre-training stage followed by supervised fine-tuning (SFT). The goal of RLHF training is to reduce the gap in naturalness between the speech LLM and human speech. To incentivize the generation of more natural speech, we curate an RLHF training dataset consisting of 9.2K samples, where the speech prompts were collected by voice actors who are instructed to choose a category from intelligence, steer ability, and emotional well being.

\textbf{RL Training.}
We adopt GRPO~\cite{guo2025deepseek} as our online reinforcement learning algorithm, which removes the critic from PPO~\cite{schulman2017proximal} and estimates the value function by averaging rewards within a group. For each input prompt, the speech LLM generates four candidate responses, which are then scored by GSRM. Specifically, GSRM evaluates each response across all sub-dimensions listed in Table~\ref{table:rubric}, together with language complexity as a semantic metric. Each sub-metric is scored on a 5-point Likert scale, and the final reward is obtained by uniformly averaging all sub-metric scores. To reduce variance in reward estimation, we run GSRM inference 20 times per response, producing eight independent judgments, and use their average as the final reward. We use a policy learning rate of $4\times10^{-7}$. The batch size is set to 320, the maximum sequence length is 8192, and the KL coefficient is set to $1\times10^{-2}$. The Speech LLM is trained for up to 2000 steps.

\textbf{Evaluation.}
For evaluation, we randomly sample 50 audio prompts from an in-house evaluation dataset. The prompts span multiple categories, including chitchat, paralinguistics, steerability, and emotional well-being. These speech prompts are collected from human volunteers, who record their own spoken questions corresponding to each category. For each prompt, both the RLHF-trained model and the base model generate speech responses. Human raters then perform A/B evaluations on the paired responses across multiple dimensions, including tone, pacing, intonation, and overall naturalness using the rubric shown below.

Each response pair is generated from the same prompt and evaluated independently by five different reviewers. To ensure consistency, we perform quality checks to verify that the pairwise preferences align with the preferences derived from comparing the absolute naturalness scores. For each response pair, we determine the final overall naturalness preference by removing outlier ratings and applying a majority vote across the reviewers. The win rate for each model is then calculated as the total number of wins (as determined by the overall preference) divided by the total number of evaluated samples.  

For the intermediate sub-dimensions, we determine model preference by directly comparing the absolute scores. A preference ranking is only requested in the event of a tie. These intermediate questions are designed to guide the rater toward an informed final decision, while remaining concise to minimize the cognitive load associated with completing multiple annotations in one sitting.

\begin{tcolorbox}[gray_box, title = {{Aesthetic Naturalness Human Evaluation Rubric }}]\footnotesize
\textbf{Rater instructions (as shown in the UI).}
\emph{``You will be presented with two different recordings of a turn
from a dialog between a human talking to an AI chat bot. Please listen
carefully to the conversation and then answer the questions below.
After listening, you are asked a series of multiple-choice questions
comparing the two responses regarding the AI chat bot (the answering
voice, not the asking voice). Do not overthink your response; choose
the option that best matches your observation.''}
\subsubsection*{Dimension 1: Pace}
\emph{Prompt.} ``Did response A/B maintain an organic, natural
\textbf{pace}? Think about the length of pauses and the usage of filler
words (e.g., `uhm', `uh') and similar nonverbal cues.''
\noindent\emph{Per-response scale (A and B, separately):}
\begin{itemize}[leftmargin=1.2em, nosep]
  \item[1] Yes, natural flow, indistinguishable from a human.
  \item[0] No, pacing was jarring or hard to listen to.
\end{itemize}
\noindent\emph{Tiebreaker (pairwise preference, only if A and B are rated the same):} \quad -1 Prefer audio A \quad 1 Prefer audio B
\subsubsection*{Dimension 2: Intonation}
\emph{Prompt.} ``How appropriate was response A/B's \textbf{intonation}
(pitch curve)? Was it jarring, or indistinguishable from a human?''
\noindent\emph{Per-response scale (A and B, separately):}
\begin{itemize}[leftmargin=1.2em, nosep]
  \item[1] Indistinguishable from a human.
  \item[0] Clearly jarring / sounds like a robot.
\end{itemize}
\noindent\emph{Tiebreaker (pairwise preference, only if A and B are rated the same):} \quad -1 Prefer audio A \quad 1 Prefer audio B
\subsubsection*{Dimension 3: Emotion and Tone}
\emph{Prompt.} ``Did response A/B's \textbf{emotion and tone} fit the
situation and context of the dialogue?''
\noindent\emph{Clarification shown.} ``For example, the assistant should
match the user’s perceived emotional state; it should not be too
cheerful, energetic, lethargic, or disrespectful given the user’s
content and tone.''
\noindent\emph{Per-response scale (A and B, separately):}
\begin{itemize}[leftmargin=1.2em, nosep]
  \item[1] Yes, appropriate (``nailed it'').
  \item[0] Not appropriate (e.g., too cheerful, bored, or disrespectful).
\end{itemize}
\noindent\emph{Tiebreaker (pairwise preference, only if A and B are rated the same):} \quad -1 Prefer audio A \quad 1 Prefer audio B
\subsubsection*{Dimension 4: Overall Naturalness and Preference}
\textbf{Pairwise preference (A vs.\ B).}
\emph{Prompt.} ``Which response appears more natural / human-like?''
\noindent\emph{Scale:}
\begin{itemize}[leftmargin=1.2em, nosep]
  \item[1] Strongly prefer audio A.
  \item[2] Slightly prefer audio A.
  \item[3] No clear preference (undecided).
  \item[4] Slightly prefer audio B.
  \item[5] Strongly prefer audio B.
  \item[X] Technical difficulty.
  \item[X] Skip.
\end{itemize}
\textbf{Absolute naturalness for each response.}
\emph{Prompt.} ``Was the response for audio A/B natural / human-like?''
\noindent\emph{Per-response scale (A and B, separately):}
\begin{itemize}[leftmargin=1.2em, nosep]
  \item[5] Indistinguishable from a human, even for a trained ear.
  \item[4] Sounds like a human for most listeners.
  \item[3] Somewhat natural, but sometimes unnatural.
  \item[2] Barely human; could tell it is a robot.
  \item[1] Clearly recognizable as a robot.
  \item[X] Technical difficulty.
  \item[X] Skip.
\end{itemize}
\textbf{Free-text rationale.}
For each response, raters provide a brief explanation of their choice,
e.g., commenting on over-acting, mismatch between tone and context,
unnatural pauses, excessive sighs, or other artifacts.
\end{tcolorbox}

\section{Additional Results}
\subsection{Correlation Between Sub-Metrics and Human-Likeness.} \label{sec:data_analysis}

\begin{figure}[h]
    \centering
    \begin{tabular}{rl}
    \subfloat[ConvTTS Dataset]{%
        \includegraphics[width=0.5\textwidth]{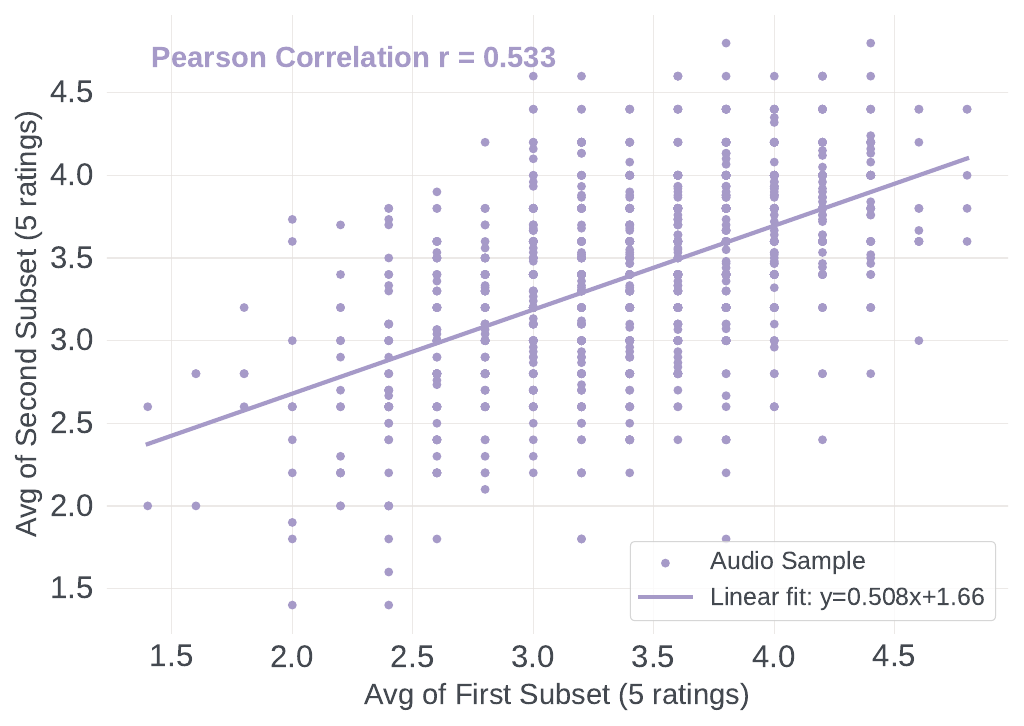}
    } &
    \subfloat[FDX-Conv Dataset]{%
        \includegraphics[width=0.5\textwidth]{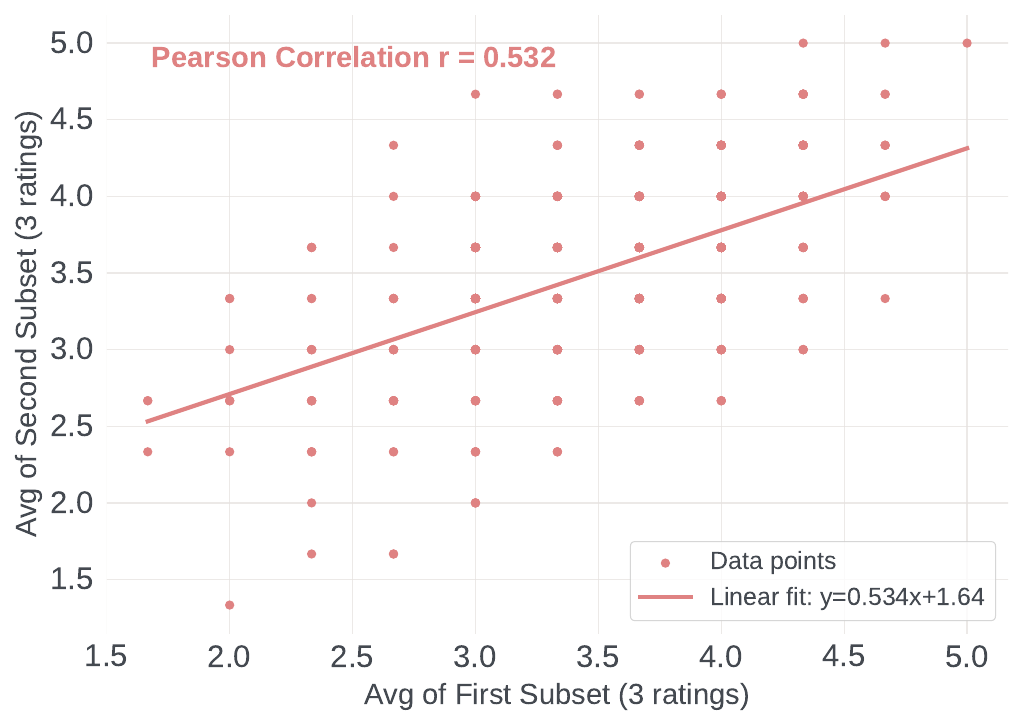}
    }
    \end{tabular}
\caption{\textbf{Inter-Rater Consistency of Human Likeness Ratings.} }
\label{fig:human_pcc}
\vspace{-1em}
\end{figure}

\begin{figure*}[h]
    \centering
    \includegraphics[width=1.0\textwidth]{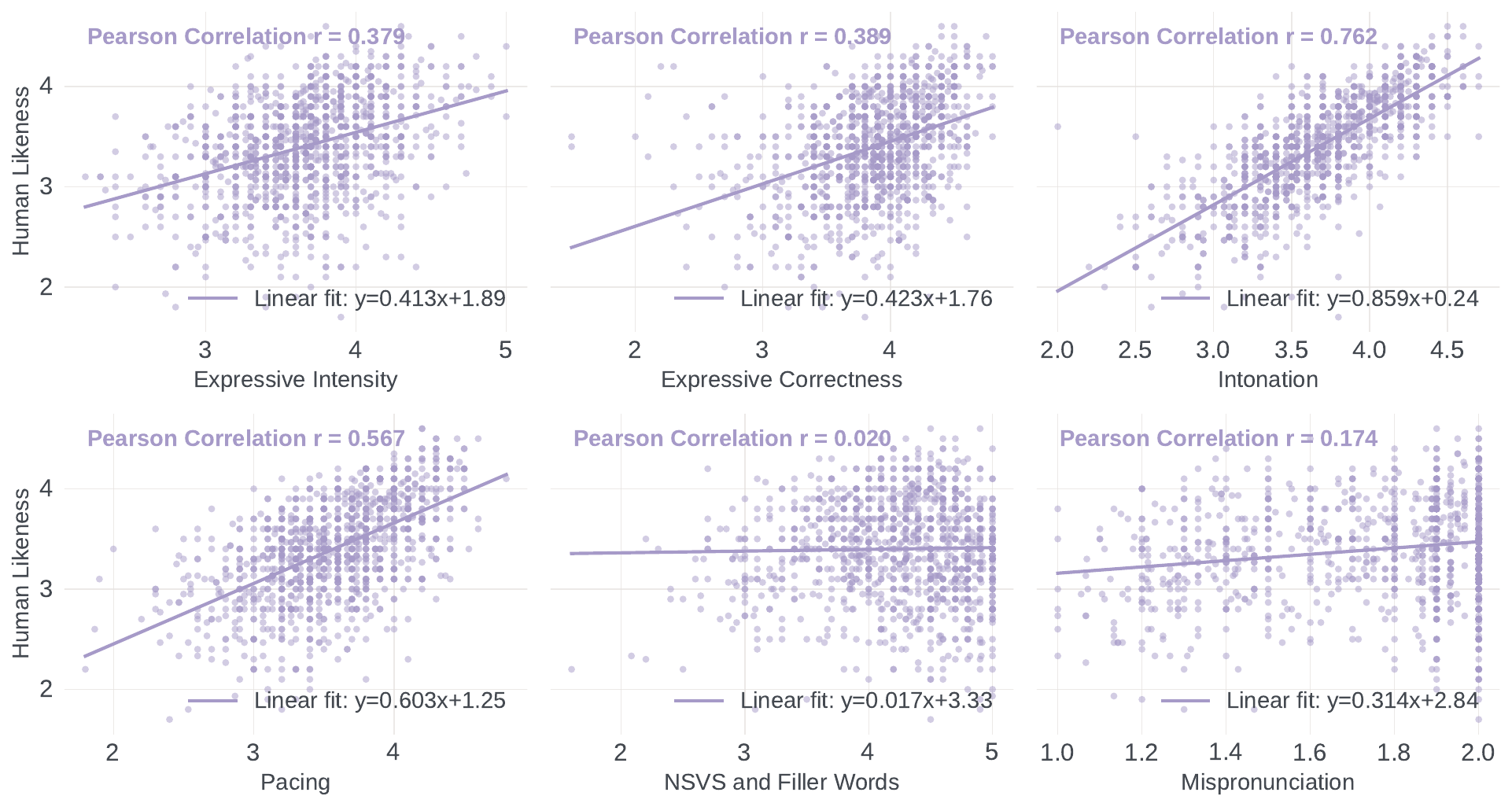}
\caption{\textbf{Correlation Between Sub-Metrics and Human-Likeness Ratings} }
\label{fig:submetric_pcc}
\end{figure*}
We further analyze how individual sub-metrics relate to the overall human-likeness score. For each sub-metric, we compute the Pearson correlation coefficient between the sub-metric ratings and the corresponding human-likeness ratings. The results shown in Figure~\ref{fig:submetric_pcc} indicate that \emph{intonation} and \emph{pacing} exhibit strong correlations with human-likeness, while \emph{expressive intensity} and \emph{expressive correctness} show moderate correlations. In contrast, \emph{non-speech vocalizations and fillers} and \emph{mispronunciation} exhibit relatively weak correlations with the overall human-likeness score.

\subsection{Training GSRM via Reinforcement Learning} \label{sec:gsrm_rl}
GSRM performs a speech-in, text-out reasoning task. Since reinforcement learning has been shown to be effective for improving reasoning capabilities in large language models, we investigate whether reinforcement learning can further enhance GSRM’s performance on speech naturalness prediction. A common training paradigm for reasoning-oriented LLMs consists of a small-scale supervised fine-tuning (SFT) warm-up stage followed by large-scale reinforcement learning~\cite{guo2025deepseek, shen2025satori}. We follow this paradigm by first fine-tuning Qwen2.5-Omni-7B with 500 supervised samples. We then apply reinforcement learning using GRPO~\cite{guo2025deepseek} on the remaining 4,079 samples.

A key challenge in reinforcement learning is reward design. We consider two reward modeling strategies. The first uses the negative $\ell_2$ distance between the predicted score $\hat{y}$ and the ground-truth rating $y^*$, defined as $-\lVert y^* - \hat{y} \rVert$. The second uses an RBF kernel to measure similarity between $\hat{y}$ and $y^*$, defined as $\exp\!\left(-\frac{(y^* - \hat{y})^2}{2\sigma^2}\right)$. Models trained with these two reward formulations are evaluated on the out-of-domain (OOD) test set and compared against a GSRM trained using full supervised fine-tuning on all 4,579 samples.

The comparison results are shown in Table~\ref{table:gsrm-rl-results}. Notably, when trained with the $\ell_2$-distance reward, the model collapses to predicting nearly constant scores across different inputs. This behavior indicates reward hacking: rather than accurately matching human ratings, the model converges to a conservative prediction strategy that minimizes expected penalty. In contrast, the nonlinear RBF-kernel reward mitigates this issue and yields modest improvements over the warm-up SFT model. However, the RL-trained models still underperform the model trained with full supervised fine-tuning.

Based on these observations, we hypothesize that the base model Qwen2.5-Omni-7B lacks sufficient prior knowledge for analyzing paralinguistic cues and judging speech naturalness. Supervised fine-tuning therefore plays a critical role as a knowledge injection stage, enabling the model to acquire the necessary skills for speech evaluation. While incorporating reinforcement learning alone is insufficient in this case, a promising direction for future work is to introduce substantial amounts of speech judgment data during pre-training or mid-training stages, which we leave for future investigation.

\begin{table}[!t]
\centering
\caption{\textbf{OOD Performance of GSRM Trained with Reinforcement Learning.} Comparison of supervised fine-tuning and reinforcement learning strategies for training GSRM, evaluated on the out-of-domain test set.}
\setlength{\tabcolsep}{3.5pt}
\begin{tabular}{l c c c}
\toprule[1.5pt]
\textbf{Method} 
& {\textbf{Pearson} $\uparrow$}
& {\textbf{Spearman} $\uparrow$}
& {\textbf{MSE} $\downarrow$} \\
\midrule
GSRM + Full SFT
& 0.465 & 0.427 & 0.230  \\
GSRM + Warm-up SFT
& 0.249 & 0.248 & 0.275  \\
GSRM + RL (L2 Distance) 
& 0.012 & 0.015 & 0.435  \\
GSRM + RL (RBF Kernel) 
& 0.270 & 0.259 & 0.272  \\
\bottomrule
\end{tabular}
\label{table:gsrm-rl-results}
\end{table}

\subsection{Results of GSRM for Speech Semantic Judge}
\label{sec: gtrm results}

\begin{table}[thb]
\centering
\caption{\textbf{Main Results of GSRM for Speech Semantic Judge.} Pearson correlation coefficients between ground-truth scores and different scoring strategies for speech semantic reward model trained with and without CoT reasoning, evaluated in-distribution (ID) and out-of-distribution (OOD).}
\label{tab: gtrm results}
\begin{tabular}{lccc|ccc}
\toprule
& \multicolumn{3}{c|}{\textbf{ID}} & \multicolumn{3}{c}{\textbf{OOD}} \\
\cmidrule(lr){2-4} \cmidrule(lr){5-7}
\textbf{Setup} & Mean & Median & Rounded Mean & Mean & Median & Rounded Mean \\
\midrule
without CoT reasoning
  & 0.803 & 0.810 & 0.810
  & 0.706 & 0.677 & 0.659 \\
with CoT reasoning
  & \textbf{0.886} & 0.874 & 0.884
  & 0.834 & 0.835 & \textbf{0.837} \\
\bottomrule
\end{tabular}
\end{table}

As discussed in Section \ref{sec: gsrm training}, we trained two semantic reward models, our proposed model: a generative reward model with CoT reasoning and a baseline without any CoT reasoning that only predicts the naturalness score.  We evaluate these two models on an in-domain evaluation set of 100 samples and an OOD dataset of 130 samples. We extract a final score for each dialogue using standard nucleus sampling with temperature of 1.0 and top-$p=0.6$. For each dialogue, we sample 10 independent predictions and then apply an aggregation rule to obtain a single scalar score. As shown in Table~\ref{tab: gtrm results}, our proposed method achieves the strongest performance, with superior generalization on OOD data. After aggregating the 10 predictions into a single score we compute the pearson correlation coefficient against the ground-turth score. Among aggregation strategies, the rounded mean works particularly well, similar to the audio naturalness GSRM discussed in the main text. The results also highlight the benefits of training with chain-of-thought (CoT) reasoning: it not only improves score prediction on in-domain data but also substantially enhances OOD generalization, addressing a common failure mode of classical regression-based reward modeling.

\begin{figure}[!thb]
  \centering
  \subfloat[without CoT reasoning]{
    \includegraphics[width=0.48\linewidth]{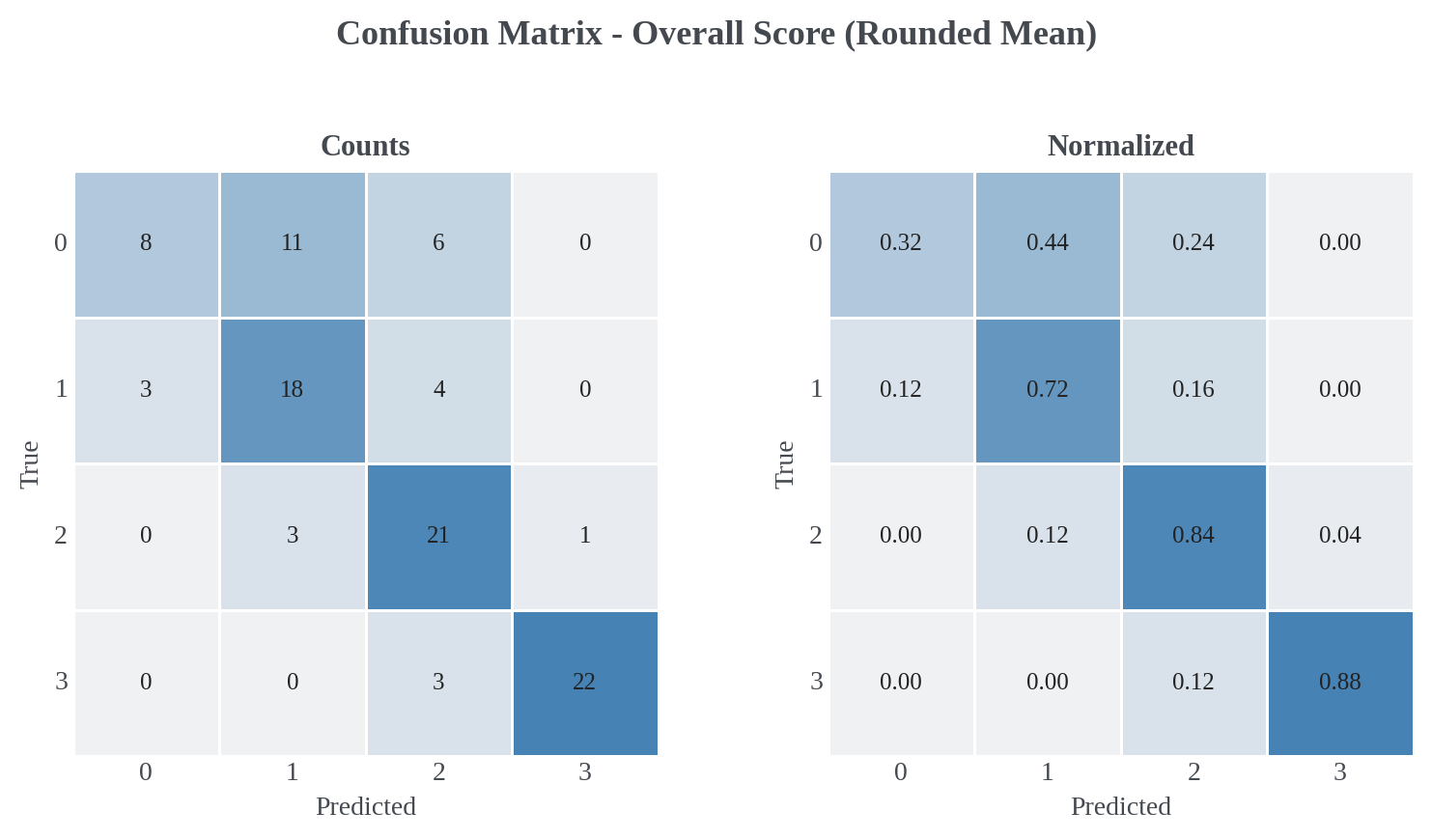}
  }
  \hfill
  \subfloat[with CoT reasoning]{
    \includegraphics[width=0.48\linewidth]{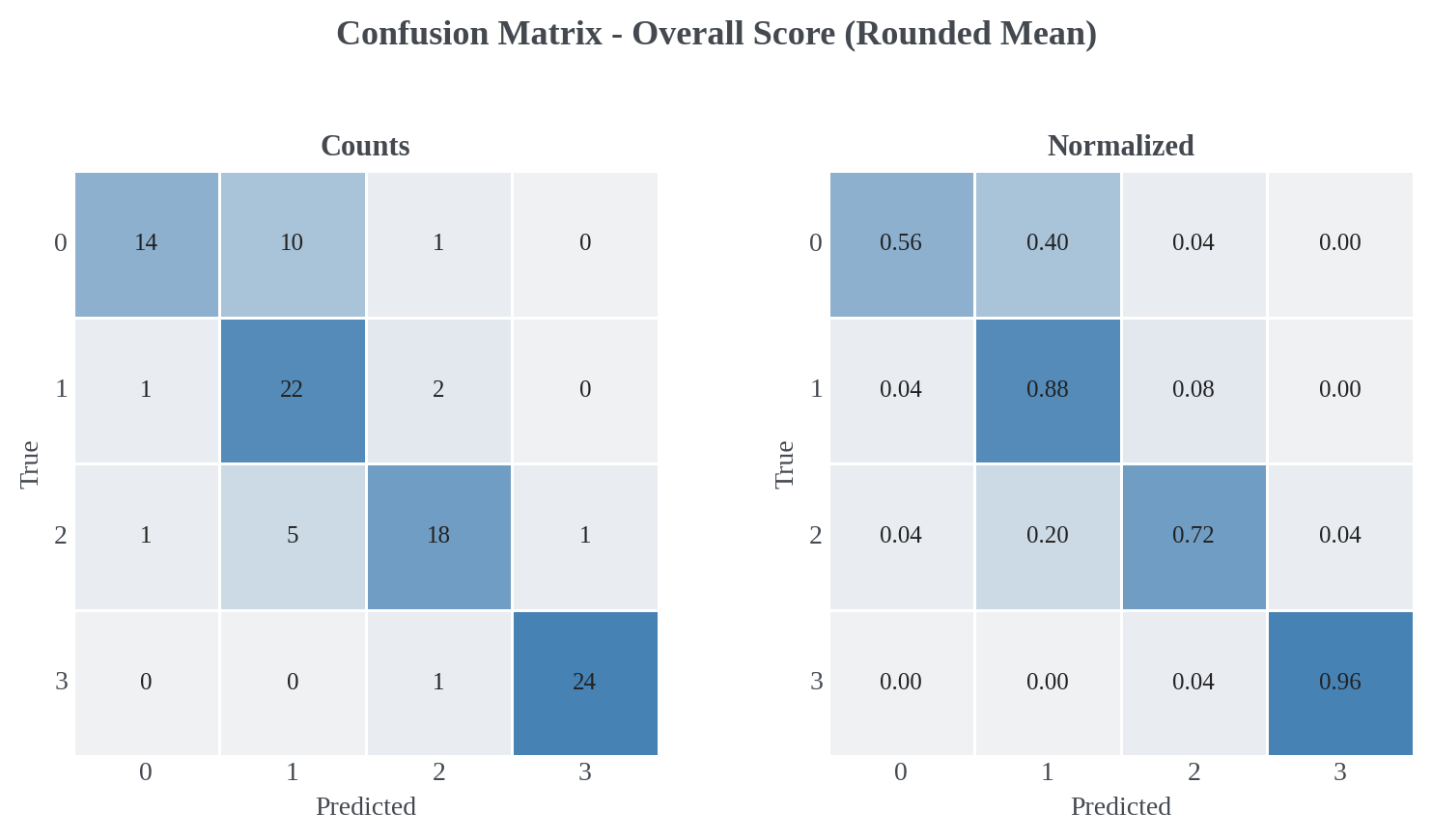}
  }
  \caption{\textbf{Confusion matrix of predictions from models evaluated on in-domain eval data.}}
  \label{fig: confusion matrix indomain}
\end{figure}

\begin{figure}[!thb]
  \centering
  \subfloat[without CoT reasoning]{
    \includegraphics[width=0.48\linewidth]{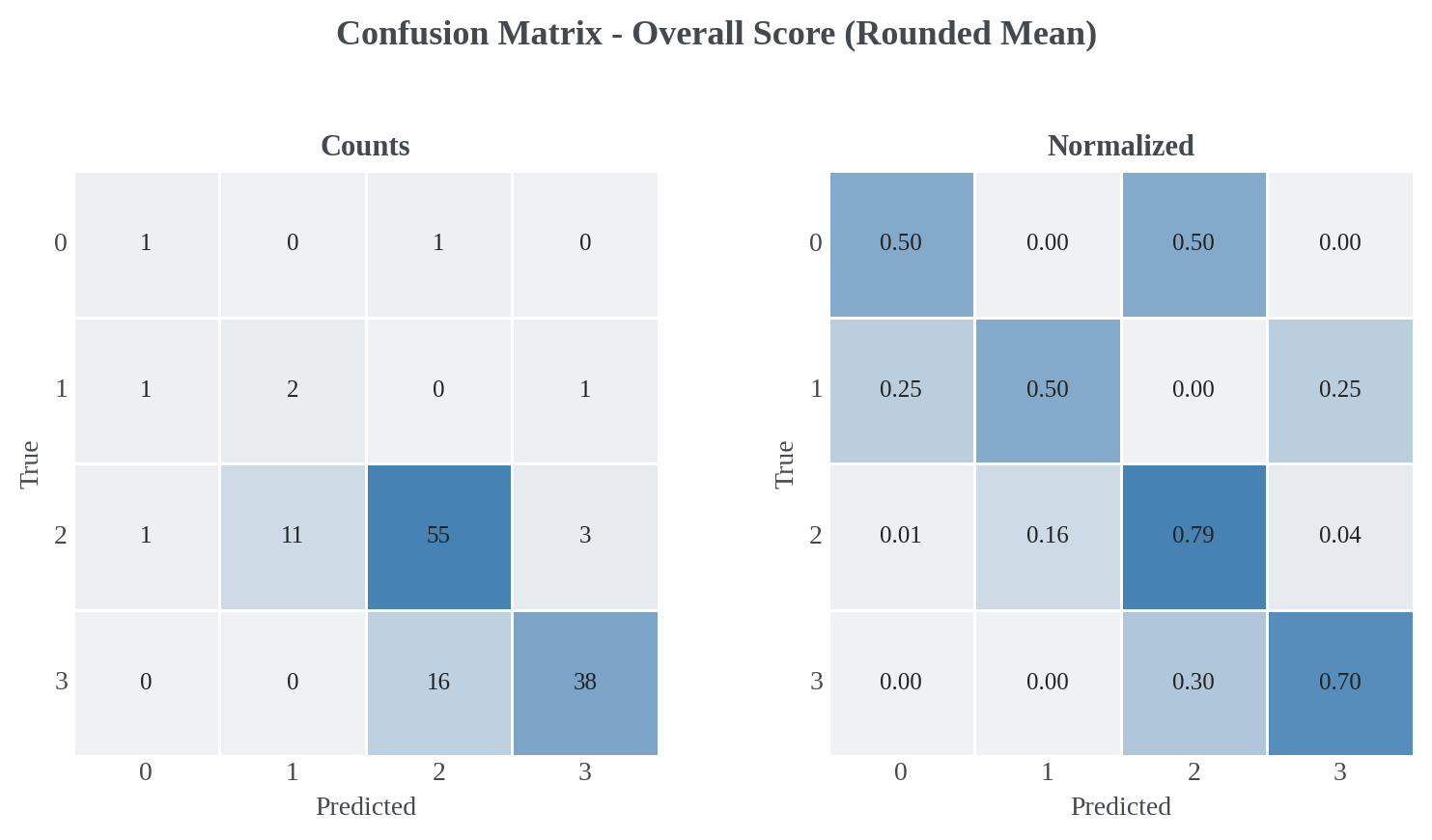}
  }
  \hfill
  \subfloat[with CoT reasoning]{
    \includegraphics[width=0.48\linewidth]{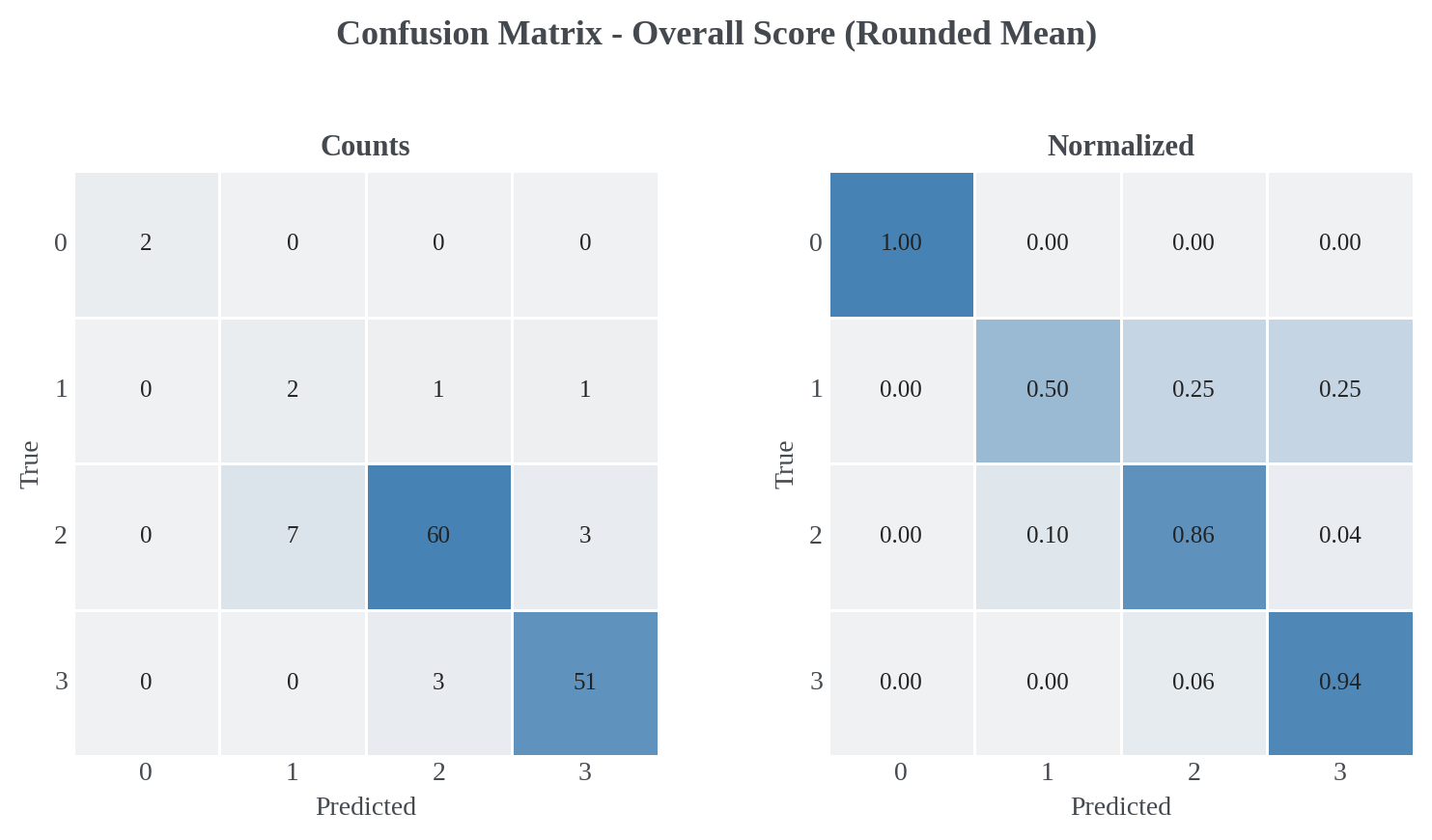}
  }
  \caption{\textbf{Confusion matrix of predictions from models evaluated on OOD eval data.}}
  \label{fig: confusion matrix ood}
\end{figure}

\begin{figure}[!thb]
  \centering
  \subfloat[Language complexity confusion matrix]{
    \includegraphics[width=0.3\linewidth]{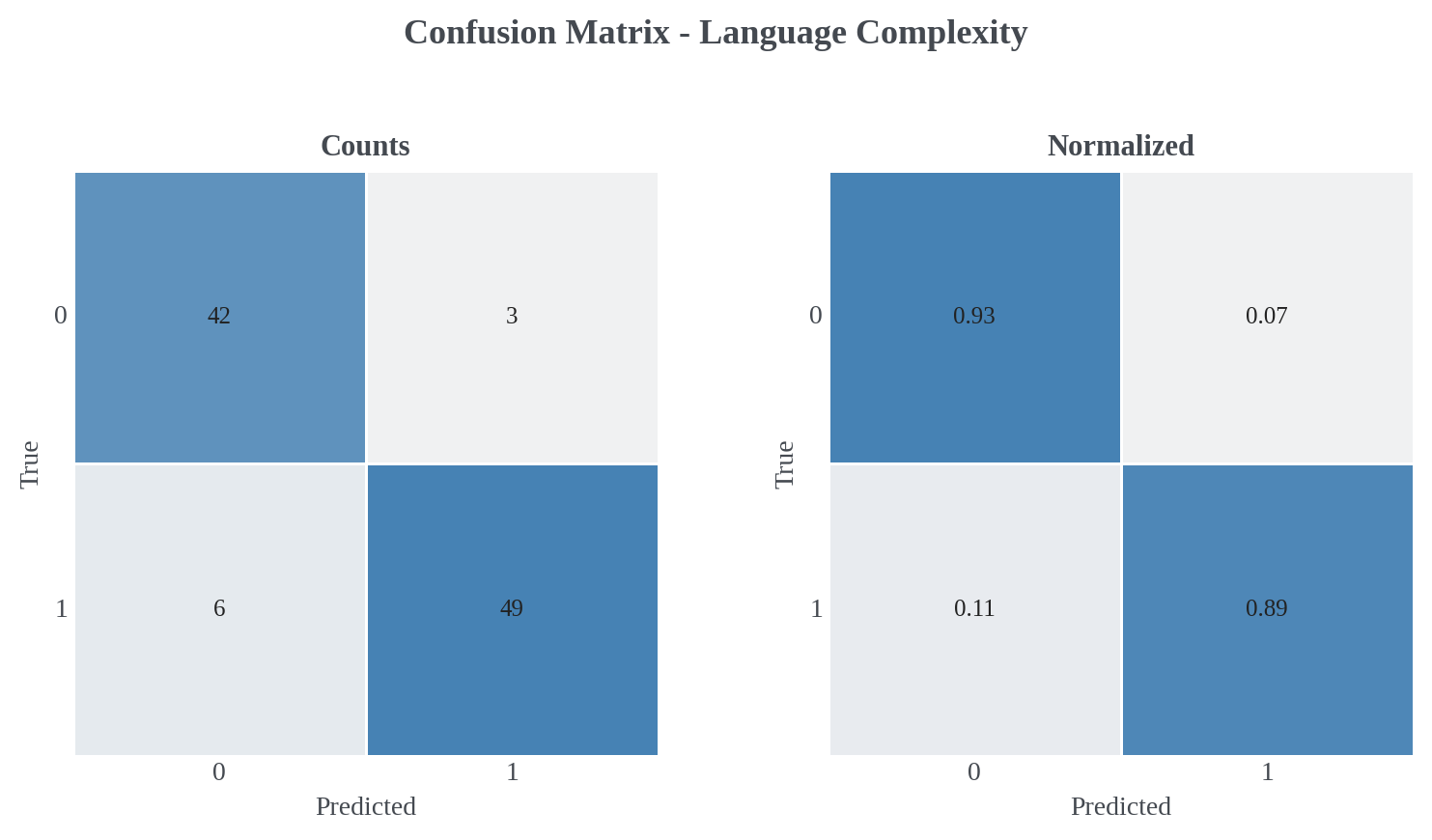}
  }
  \hfill
  \subfloat[Spontaneity confusion matrix]{
    \includegraphics[width=0.3\linewidth]{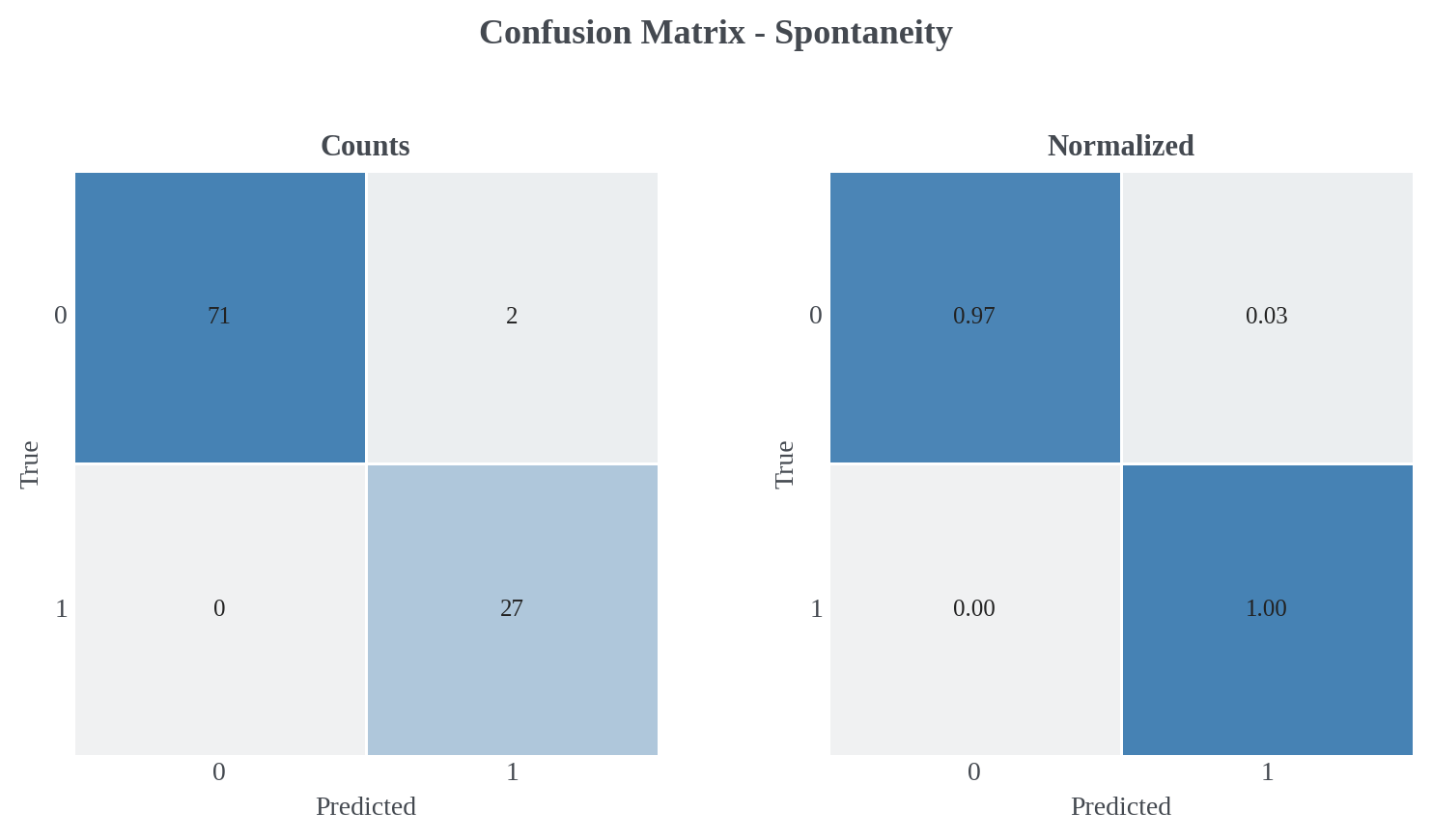}
  }
  \hfill
  \subfloat[Contextual awareness confusion matrix]{
    \includegraphics[width=0.3\linewidth]{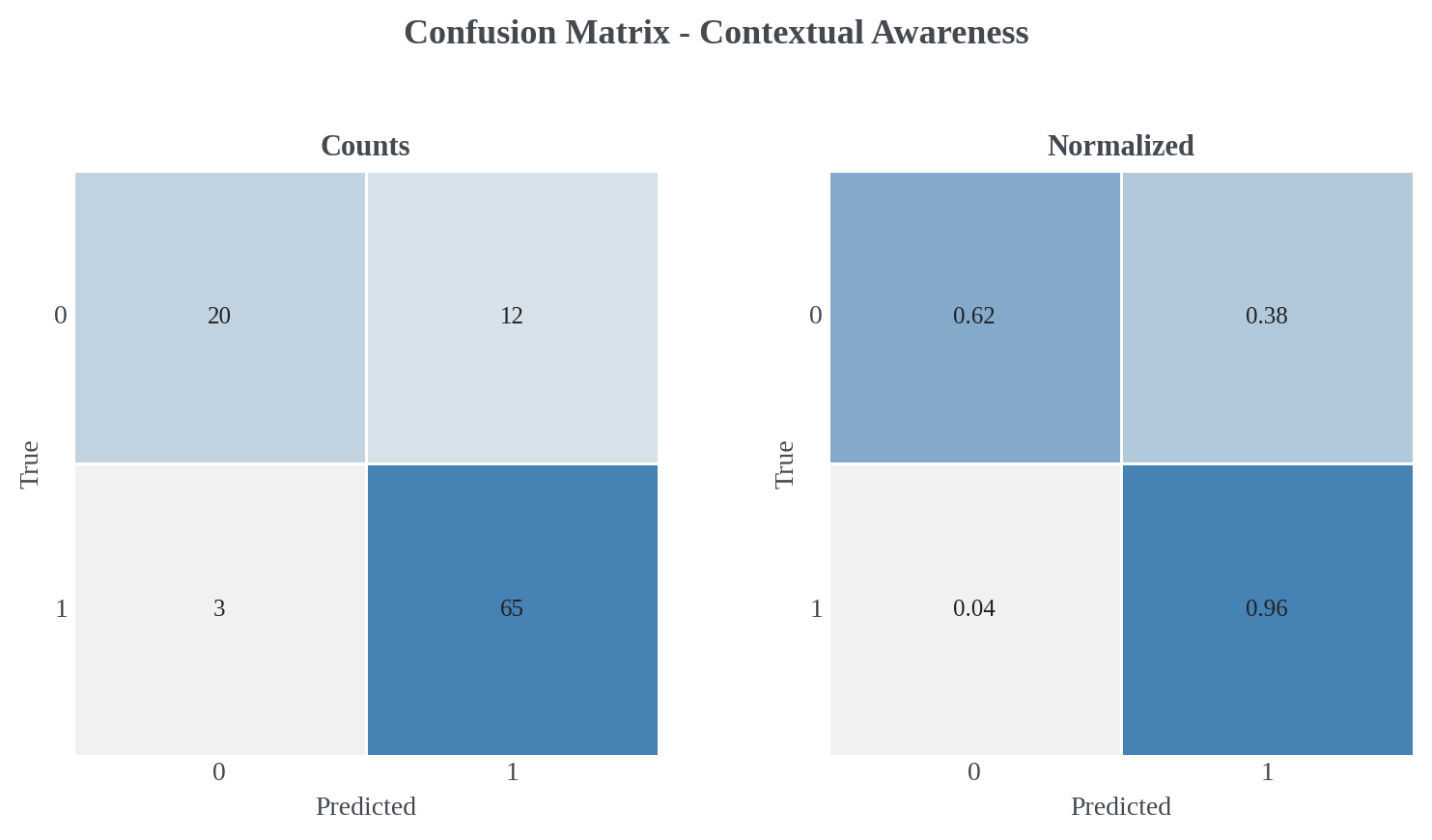}
  }
  \caption{\textbf{Confusion matrices on predictions of sub-dimension scores from model trained with CoT reasoning.}}
  \label{fig: subdimension confusion matrices}
\end{figure}

The confusion matrices presented in Figures~\ref{fig: confusion matrix indomain} and~\ref{fig: confusion matrix ood} demonstrate that incorporating CoT reasoning enables the reward model to achieve substantially higher accuracy, particularly for samples at the extremes of the naturalness spectrum (i.e., scores of 0 and 3). Notably, this improvement persists even on out-of-distribution (OOD) data, despite the limited number of samples with low naturalness scores.

Finally, Figure~\ref{fig: subdimension confusion matrices} illustrates the performance of our model, trained with CoT reasoning, in predicting sub-dimension scores for language complexity, spontaneity, and contextual awareness. The model achieves an agreement rate exceeding 85\% for both language complexity and spontaneity. However, its performance in identifying a lack of contextual awareness, particularly in lower-quality samples, remains an area for improvement. We anticipate that incorporating additional negative samples targeting this sub-dimension in future training iterations will further enhance the model’s predictive capabilities.


\end{document}